\documentclass[12pt]{article}

\usepackage{graphicx}
\usepackage[utf8]{inputenc}
\usepackage{amsmath}
\usepackage{bm}
\usepackage{fullpage}
\usepackage{indentfirst}
\usepackage[colorinlistoftodos]{todonotes}
\usepackage{scicite}

\usepackage{times}

\newcommand{\meff}{M_{\text{eff}}}
\newcommand{\ms}{M_{\text{s}}}

\newcommand{\NM}{\hat{m}}

\newenvironment{sciabstract}{%
\begin{quote} \bf}
{\end{quote}}

\title{Sagnac interferometry for high-sensitivity optical measurements of spin-orbit torque} 

\author
{Saba Karimeddiny,$^{1\dagger}$ Thow Min Jerald Cham,$^{1\dagger}$ Orion Smedley,$^{1}$\\
Daniel C. Ralph,$^{1,2\ast}$ Yunqiu Kelly Luo$^{1,2,3\dagger\ast}$\\
\\
\normalsize{$^{1}$Cornell University, Ithaca, NY 14850, USA}\\
\normalsize{$^{2}$Cornell Kavli Institute at Cornell, Ithaca, NY 14853, USA}\\
\normalsize{$^{3}$Department of Physics and Astronomy, University of Southern California,}\\
\normalsize{Los Angeles, CA 90089, USA}\\
\normalsize{$^\ast$Corresponding authors. E-mail: dcr14@cornell.edu, kelly.y.luo@usc.edu}\\
\normalsize{$^\dagger$These authors contributed equally to this work.}\\
}

\date{}
\begin{document} 
\baselineskip20pt
\maketitle 
\begin{sciabstract}
    Sagnac interferometry can provide a significant improvement in signal-to-noise ratio compared to conventional magnetic imaging based on the magneto-optical Kerr effect (MOKE).  We show that this improvement is sufficient to allow quantitative measurements of current-induced magnetic deflections due to spin-orbit torque even in thin-film magnetic samples with perpendicular magnetic anisotropy for which the Kerr rotation is second-order in the magnetic deflection. Sagnac interfermometry can also be applied beneficially for samples with in-plane anisotropy, for which the Kerr rotation is first order in the deflection angle. Optical measurements based on Sagnac interferometry can therefore provide a cross-check on electrical techniques for measuring spin-orbit torque. Different electrical techniques commonly give quantitatively inconsistent results, so that Sagnac interferometry can help to identify which techniques are affected by unidentified artifacts. 
\end{sciabstract}



\section*{Introduction}

Spin-orbit torques (SOTs) \cite{Miron2011,Liu2012} are of interest for achieving efficient manipulation of magnetization for low-power non-volatile magnetic memory technologies. SOTs are produced when a charge current is applied through a channel with strong spin-orbit coupling, giving rise to a transverse spin current. This spin current can exert a spin-transfer torque on an adjacent ferromagnet (FM), allowing for low-power electrical control of its magnetic orientation.  Accurate quantitative measurements of the efficiency of spin-orbit torques are important for understanding the microscopic mechanisms of the torque and for optimizing materials for applications. The work-horse techniques for this purpose have been electrical measurements of current-induced magnetic reorientation with readout based on the magnetoresistance properties of the samples \cite{Liu2012,pi2010,Liu2012PRL,Garelo2013,Hayashi2014,woo2014enhanced,Pai2016,Fan2016SOTTI,avci2017current,baek2018spin,mendil2019current,alghamdi2019highly,wang2019current,gupta2020manipulation,yanez2021spin,cogulu2022quantifying,cheng2022third,Gibbons2022}, but these have some shortcomings. One must be careful to separate thermoelectric voltages from the torque signals \cite{Avci2014,Roschewsky2019}, and even when performed carefully, different electrical techniques can often produce quantitatively inconsistent measurements, indicating that some may be affected by artifacts which are not yet understood \cite{Karimeddiny2020,Karimeddiny2021,cham2022,Fei2020TMDFM,Fei2020spinhall}. Furthermore, in cases when one wishes to measure spin-orbit torques acting on insulating magnetic layers,  electrical measurements provide much lower signal levels compared to metallic magnets due to decreased magnetoresistance. Optical techniques based on the magneto-optical Kerr effect (MOKE) have been introduced as an alternative to quantify spin-orbit torques \cite{Fan2014,Fan2016,Montazeri2015}, 
but in previous studies the sensitivity of MOKE measurements has been insufficient to measure current-induced small-angle magnetic deflection in samples with perpendicular magnetic anisotropy (PMA) -- the most-direct approach for quantifying the torque in the class of samples of primary interest for high-density memory applications.   

In this work, we demonstrate improved optical detection of SOTs by using a fiber Sagnac interferometer to measure current-induced small-angle magnetic tilting. Unlike conventional MOKE measurements that rely on a single laser beam, Sagnac interferometry uses the modulated phase difference of two coherent beams that travel along overlapping paths, and are incident on the sample with opposite helicities. By detecting the resulting light intensity of the interfering beams, we achieve signal-to-noise ratios at least 50 - 100 times greater than conventional MOKE  performed on a PMA metallic thin film (SI Section V). This allows us to perform accurate, highly-sensitive measurements of the spin-orbit-torque vectors in both PMA samples and in-plane anisotropy samples, based on direct optical detection of magnetization deflection in the out-of-plane (OOP) direction. 

\section*{Results}
\subsection*{Principles of Sagnac interferometry}
Our Sagnac interferometer consists of free-space optics and a 15-meter-long single-mode polarization-maintaining fiber in a compact table-top setup. As shown in Fig.~\ref{beamline}, two spatially-overlapping, orthogonal linearly-polarized beams travel inside the fiber along its fast and slow axes. Both beams pass through a quarter-wave plate to become left- and right-circularly polarized, reflect from the sample, and then pass back through the quarter wave plate to re-enter the fiber, thereby returning via the opposite fiber axis. The two beams therefore traverse the same optical path (in opposite directions) with phase and amplitude differences determined by the differences in reflection of left and right circularly-polarized light from the sample.  To measure this phase difference (i.e., 2$\theta_k$, where $\theta_k$ is the Kerr rotation angle of the sample) one can modulate the phase difference of the two beams using an electro-optic modulator (EOM). When the EOM phase modulation frequency $\omega$ matches the total optical path $\tau$ ($\omega=\pi/\tau = 2\pi$ (3.347 MHz) for our apparatus), the Kerr rotation can be quantified as
\begin{align}
\begin{split}
    \theta_k = -\frac{1}{2}\arctan\left[\frac{V_\text{APD}^{\omega}J_2(2\phi_m)}{V_\text{APD}^{2\omega}J_1(2\phi_m)}\right],
\end{split}
\label{Sagnac Kerr}
\end{align}
where $V_\text{APD}^{1\omega}$ and $V_\text{APD}^{2\omega}$ are the first and second harmonic intensity signals from the interferometer, $\phi_m$ is the EOM phase modulation depth between the fast and slow axes, and $J_\text{1(2)}$ are the Bessel functions. Details of this derivation and more information about the Sagnac apparatus and its operation are provided in the supporting information.

For demonstration purposes, we will describe measurements on two thickness series of Pt(4 nm)/Co(0.86 - 1.24 nm)/MgO(1.9 nm)/Ta(2 nm) and Pt(4 nm)/Co(1.39 - 2.08 nm)/MgO(1.9 nm)/Ta(2 nm) samples in which the Co layer is deposited as a wedge to provide a range of thicknesses on the same wafer. The samples are made by sputtering on a high-resistivity Si/SiO$_2$ wafer with a 1.5 nm Ta seed layer. They are patterned into 20 \textmu m $\times$ 80 \textmu m Hall bars with 6 \textmu m side contacts by photolithography and ion milling. The Pt resistivity for each series are 40 \textmu ohms cm and 54  \textmu ohms cm respectively (see SI Section VI. B for details). All measurements are performed at room temperature.

Magnetic hysteresis loops can be obtained by measuring $\theta_k$ while sweeping an external magnetic field.  The lower-left inset in Fig.~\ref{beamline} shows a hysteresis loop as a function of out-of-plane magnetic field for a Pt(4 nm)/Co(1.15 nm)/MgO bilayer sample with perpendicular magnetic anisotropy.  We achieve a sensitivity in measuring $\theta_k$
of better than 5 \textmu Rad/$\sqrt{\text{Hz}}$ for an average laser power of 1 \textmu W at the avalanche photodetector (APD in Fig.~\ref{beamline}), sufficient so that the noise level is not easily visible in Fig.~\ref{beamline}. While conventional MOKE can achieve comparable sensitivity using external modulation of magnetic field, electric field, or current \cite{Kato2004,Lee2017}, these methods are not applicable for measuring hysteresis curves of ferromagnets. 

The Sagnac signal is sensitive only to the out-of-plane component $m_z$ of the magnetization unit vector, with no measurable dependence on the in-plane components. For linearly-polarized light incident on the sample in the normal direction, the quadratic MOKE effect does allow a second-order dependence on the in-plane magnetization components in that the total Kerr rotation can have the form \cite{Fan2016} 

\begin{align}
    \theta_k = \kappa m_z + \beta_Q m_x m_y
    \label{Qmoke}
\end{align}
where $\kappa$ is a material-specific constant of proportionality relating the out-of-plane net magnetization to $\theta_k$, $\beta_Q$ is the quadratic MOKE coupling parameter, and $m_x$ and $m_y$ are  defined such that $x$ lies along the plane of light polarization. However, we calculate that the contribution of quadratic MOKE to the Sagnac signal is approximately a factor of 10$^{-5}$ smaller than the $\kappa m_z$ contribution (see SI Section III).  Furthermore, the quadratic MOKE contribution to the Sagnac signal should introduce a dependence $\propto \sin(2\phi)$, where $\phi_\text{}$ is the angle between the in-plane magnetization and a reference plane of light polarization. No such dependence is measurable in Sagnac measurements if we apply in-plane field of fixed magnitude and then rotate $\phi_\text{}$ (see Supplemental Fig.~\ref{quadmoke}). Based on both calculations and measurements we therefore conclude that the Sagnac signal depends measurably only on $m_z$. The absence of dependence on the in-plane magnetization components simplifies the Sagnac measurements of spin-orbit torque relative to, e.g., electrical measurements of the second harmonic Hall effect \cite{Hayashi2014}, for which planar Hall signals are assumed to affect the signals in addition to the anomalous Hall effect.

\vspace{4mm}
\subsection*{Using Sagnac interferometry to measure spin-orbit torques}
We measure current-induced torques by applying a calibrated low-frequency AC current along the X direction ($\omega_e$ = 3.27 kHz) to the heavy metal/ferromagnet bilayers and measuring the resulting small-angle deflection of the magnetization. The deflection is detected from the Sagnac signal using a side-band demodulation technique, allowing us to simultaneously measure both the the steady-state value $\theta_k$ demodulated at the EOM frequency $\omega$, and the current-induced change $\Delta\theta_k$  at the lower side-band frequency $\omega - \omega_e$. We achieve a current-modulated Kerr rotation sensitivity of 3 \textmu Rad/$\sqrt{\text{Hz}}$, allowing us to detect small changes of $m_z$ due to current-induced torques. The AC current frequency $\omega_e$ is sufficiently low for the magnetic dynamics to be quasi steady-state. Therefore by balancing torques within the Landau-Lifshitz-Gilbert-Slonczewski equation \cite{Ralph2008} in steady state, the current-induced damping-like and field-like effective torques (per unit magnetization) $\tau^0_\text{DL}$ and $\tau^0_\text{FL}$ can be determined from the  deflection of the magnetic unit vector $\Delta \hat{m}$ according to 
\begin{align}
\begin{split}
\gamma \mu_0 \Delta \hat{m} \times \vec{H}_\text{eff} = \tau^0_\text{DL}\NM\times\left(\hat{\sigma}\times\NM\right) + \tau^0_\text{FL}\hat{\sigma}\times\NM
\label{LLGS}
\end{split}
\end{align}
where $\gamma=2\mu_B/\hbar$ is the gyromagnetic ratio with $\mu_B$ the Bohr magneton, and $\vec{H}_\text{eff}$ is the vector sum of the anisotropy field and any applied magnetic field. We assume here that the spin-source layer has high symmetry, so that the orientation of the current-induced spin polarization is parallel to $\hat{Y}$, i.e., in the sample plane and perpendicular to the charge current (shown in Fig.~\ref{beamline} middle inset).

\subsection*{Samples with perpendicular magnetic anisotropy}
We first consider the case of samples with perpendicular magnetic anisotropy (PMA), which is the more difficult case for optical measurements of spin-orbit torque since the measured changes in the OOP magnetization are second order in small-angle tilting from the OOP direction. In the presence of an in-plane applied magnetic field $H$ and in the absence of applied current, the equilibrium polar angle of the magnetization $\theta_0$ (measured from $z$-axis) satisfies $\sin\theta_0 = H/\meff$, where the effective magnetization $\mu_0\meff = \mu_0 \ms - 2K_\perp/\ms$
, is the out-of-plane anisotropy minus the saturation magnetization (with a positive sign for PMA samples)\cite{Hayashi2014}. Therefore, Kerr rotation associated with the magnetic-field-induced equilibrium tilt angle ($\theta_0$) is approximately
\begin{align}
    \theta_k &= \pm\kappa\left(1 - \frac{H^2}{2\meff^2}\right) \label{ACthetak},
\end{align}
where the $\pm$ corresponds to the initial out-of-plane magnetization $m_Z = \pm 1$ (see SI Section IV for details). From Eq.~(\ref{LLGS}), the current-driven effective field in the $X$ direction corresponds to the damping-like torque: $\mu_0 \Delta H_X = \mp \tau^0_\text{DL}/\gamma$. The current-induced effective field in the $Y$ direction is the sum of the field-like spin-orbit-torque contribution and the {\O}rsted field: $\mu_0 \Delta H_Y = \mu_0 H_\text{Oe} + \tau^0_\text{FL}/\gamma$. 

In order to measure the current-driven effective fields $\Delta H_X$ and $\Delta H_Y$ for samples with PMA, we apply an  in-plane magnetic field along the $X$ or $Y$-axis ($H_X$ at $\phi_H = 0$ or $H_Y$ at $\phi_H =\pi/2$, where $\phi_H$ is the angle of the in-plane field relative to the current direction) for both of the cases $m_z = \pm 1$ and perform simultaneous measurements of $\theta_k$ and $\Delta\theta_k$.  The left two panels of Fig.~\ref{PMA Sagnac} shows the results for the same PMA Pt(4 nm)/Co(1.15 nm)/MgO bilayer for which the out-of-plane hysteresis curve is shown in Fig.~\ref{beamline}, for an AC current amplitude of 15 mA corresponding to a current density in the Pt layer of 1.9 $\times 10^{7}$ A/cm$^2$. Because $\Delta H_X$ and $\Delta H_Y$ cause small oscillations of the magnetization, the current-induced Kerr rotation (derived in SI Section IV) can be approximated as
\begin{align}
    \Delta\theta_k &= \mp\kappa\left(\Delta H_X \cos\phi_H +\Delta H_Y \sin\phi_H\right) \frac{H}{\meff^2}\label{ACdeltathetak}.
\end{align}
Therefore, $\Delta H_X$ and $\Delta H_Y$  can be extracted based on equations (\ref{ACthetak}) and (\ref{ACdeltathetak}) as
\begin{align}
    \Delta H_{X} &= \frac{d\Delta \theta_k (\phi_H=0)}{dH}\left(\frac{d^2\theta_k}{dH^2}\right)^{-1} \label{HXSagnac} \\
        \Delta H_{Y} &= \frac{d\Delta \theta_k (\phi_H=\pi/2)}{dH}\left(\frac{d^2\theta_k}{dH^2}\right)^{-1}. \label{HYSagnac}
\end{align}
 For the current amplitude of 15 mA, we find $\mu_0 \Delta H_X = \mu_0 \Delta H_\text{DL} = 5.0(3)$ mT and $\mu_0 \Delta H_Y = \mu_0 \Delta H_\text{FL}$ = -0.9(2)  mT for $ m_z = +1$, and  $\mu_0 \Delta H_X = - \mu_0 \Delta H_\text{DL} = -5.1(3)$ mT and $\mu_0 \Delta H_Y = \mu_0 \Delta H_\text{FL} = -0.9(2)$ mT for $m_z = -1$. These signs are consistent with the directions of the damping-like and field-like effective fields measured by harmonic Hall and spin-torque FMR from Pt \cite{Liu2012PRL,Garelo2013,ou2016origin}.

 We can also express these results in terms of dimensionless SOT efficiencies $\xi_\text{DL}$ and $\xi_\text{FL}$:
\begin{align}
    \xi_\text{DL(FL)}= \tau^0_\text{DL(FL)}\frac{e M_s t_\text{Co}}{\mu_B J_e}\label{taudef}
\end{align}
where $J_e$ is the electric current density in the spin source layer, $M_s$ is the saturation magnetization of the FM, and $t_\text{Co}$ is the thickness of the FM Cobalt layer. (Note by this definition that $\xi_\text{FL}$ contains contributions from both the {\O}rsted torque and the field-like SOT.) For each of our samples we calibrate the the saturation magnetization per unit area $M_s t_\text{Co}$ using vibrating-sample magnetometry on 3 mm $\times$ 3 mm thin films diced from the wafer adjacent to the patterned devices (see SI Section VI. C). We calculate $J_e$ using a parallel-conduction model after determining the thickness-dependent conductivities of the different layers in the heterostructure (See SI section VI. B). 
For the most acccurate determination of the  torque efficiencies, we measure $\Delta H_X$ and $\Delta H_Y$ for a sequence of applied voltage amplitudes for $m_z = +1$ and fit to a linear dependence (Fig.~\ref{PMA Sagnac}(e)). We can then extract $\xi_\text{DL(FL)}$ based on the fitted linear slope from Eq.~(\ref{taudef}). For the PMA Pt(4 nm)/Co(1.15 nm)/MgO bilayer we find $\xi_\text{DL} = 0.132(2)$ and $\xi_\text{FL} = -0.023(2)$. We will analyze below the results for full thickness series of the Co layer.

\vspace{4mm}
\subsection*{Samples with in-plane magnetic anisotropy}
\begin{figure*}[h]
    \centering
    \includegraphics[width=\linewidth]{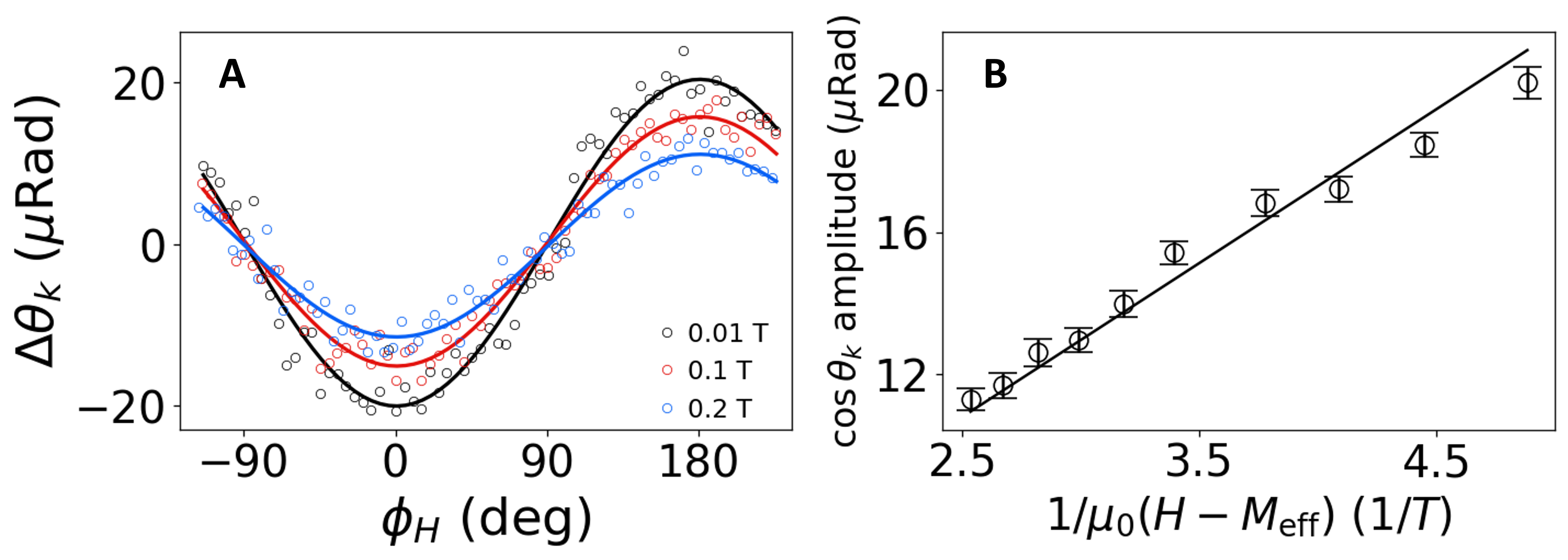}
    \caption{\textbf{Sagnac interferometry measurements of current-induced torque for a Pt(4 nm)/Co(1.42 nm)/MgO sample with in-plane magnetic anisotropy.} (\textbf{A}) $\Delta \theta_k$ as a function of in-plane magnetic field angle $\phi_H$ at 0.1, 0.15, and 0.2 T. (\textbf{B}) Amplitudes of the $\cos\phi_H$ component for different applied field magnitudes. The linear slope as a function of $1/\mu_0(H - \meff)$ allows extraction of the damping-like effective field based on Eq.~(\ref{IPtiltEq}).}
    \label{IP Sagnac}
\end{figure*}

For the case of samples with in-plane anisotropy, the current-induced changes in $m_z$ are first order in the tilting angle for out-of-plane magnetic deflections. Based on Eq.~(\ref{LLGS}), for in-plane magnetization, the damping-like torque corresponds to an out-of-plane effective field while the field-like torque gives an in-plane effective field. Therefore, our Sagnac MOKE interferometry measures only the out-of-plane magnetic deflection from the damping-like effective field, with the maximum magnitude (for $\phi_H = 0$) of $\mu_0\Delta H_\text{DL} = \tau^0_\text{DL}/\gamma$, and $\Delta\theta_k$ (derived in SI Section IV) can be expressed as

 \begin{align}
     \Delta\theta_k = &\frac{\kappa\Delta H_\text{DL} \cos\phi_H}{H - \meff}. \label{IPtiltEq}
\end{align}
Figure \ref{IP Sagnac}(a) shows $\Delta\theta_k$ as a function of of the angle of the in-plane magnetic field $\phi_H$ with constant magnitudes of magnetic field ($\mu_0 H$ = 0.1, 0.15, and 0.2 T), and a current amplitude of 8 mA for a bilayer with the composition Pt(4 nm)/Co(1.42 nm)/MgO which has in-plane magnetic anisotropy. To quantify $\Delta H_\text{DL}$, we fit the amplitude of the cos$\phi_H$ components as a function of $1/[\mu_0(H-\meff)]$ and perform a linear fit as shown in Fig.~\ref{IP Sagnac}(b). We also determine the effective magnetization $M_\text{eff}$ for each device from spin-torque ferromagnetic resonance measurements (ST-FMR) (SI Section V.D). For the device featured in Fig.~\ref{IP Sagnac}, $\mu_0M_\text{eff}$ =0.195 T, and the final result of the measurement is $\mu_0 \Delta H_\text{DL}$ = 3.0(1) mT, corresponding to $\xi_\text{DL}$ = 0.10(1).  

\vspace{4mm}
\subsection*{Results for samples over the full thickness range} 
The results of the Sagnac-interferometer measurements of SOT efficiencies for the full range of thicknesses for the Pt(4 nm)/Co(0.85 - 2.1 nm)/MgO are shown in Fig.~\ref{torques}. By varying the Co thickness, competition between the in-plane shape anisotropy and interface perpendicular magnetic anisotropy gives rise to different values of $M_\text{eff}$ (plotted in SI Fig.~\ref{Meff_STFMR}). We observe at most only a weak dependence of $\xi_\text{DL}$ on the Co layer thickness (Fig.~\ref{torques} (a) and (b)). This is expected as long as the Co layer is sufficiently thick for full absorption of the transverse component of the incoming spin current, and qualitatively consistent with previous electrical measurements \cite{Pai2015}. The values of $\xi_\text{DL}$ obtained by the Sagnac measurements on PMA and in-plane samples are consistent, which is often not the case for electrically-based second-harmonic Hall measurements of SOT \cite{karimeddiny2021sagnac}. This value that we find for the damping-like SOT efficiency is also in quantitative agreement with spin-torque ferromagnetic resonance measurements with similar Pt resistivity \cite{Zhu_2019,Karimeddiny2020,Liu2011}. Because the Sagnac interferometry is sensitive only to out-of-plane magnetic deflections, we obtain measurements of the current-induced field-like torque only for the PMA samples, in which case the field-like torque efficiency $\xi_\text{FL}$ is considerably smaller than $\xi_\text{DL}$ as shown in Fig.~\ref{torques}(b). The estimated Oersted torque is of similar amplitude as indicated in pink line in Fig.~\ref{torques}(b). This indicates that the field-like spin-orbit torque is at most a small contribution. 

\section*{Discussion}
We have shown that Sagnac interferometry provides a sufficient improvement in the signal-to-noise ratio compared to conventional MOKE to enable for the first time optical measurements of spin-orbit torque efficiencies even for thin-film magnetic samples with out-of-plane magnetic anisotropy for which the Kerr signal is second-order in the magnetic deflection angle.  The Sagnac technique also allows optical measurements of the damping-like component of spin-orbit torque for samples with in-plane magnetic anisotropy, the component of torque that causes out-of-plane magnetic deflections in this geometry.  (Measurements for the in-plane geometry have also been performed previously using conventional MOKE \cite{Fan2014,Fan2016,Montazeri2015}.)  Optical measurements provide the capability to perform quantitative studies of spin-orbit torque in samples for which magnetoresistance signals are small (e.g., insulating magnetic layers).  They can also provide an important cross-check on electrical measurements of spin-orbit torque, to identify cases in which the electrical measurements are affected by unknown artifacts.  In our Pt/Co wedge series samples, we find that the Sagnac measurements of the damping-like spin-orbit torque efficiency are in reasonable quantitative agreement throughout the thickness series for the magnetic layer, for samples with both perpendicular magnetic anisotropy (PMA) and in-plane anisotropy. These values are also in good agreement with spin-torque ferromagnetic resonance measurements with similar Pt resistivity \cite{Zhu_2019,Karimeddiny2020,Liu2011}. However, as we have noted in a separate arXiv posting, low-frequency second-harmonic electrical measurements for the PMA samples yield results that are inconsistent with both the Sagnac measurements and the ST-FMR results on the in-plane samples.  The Sagnac results therefore provide confirmation of the ST-FMR values and reason to question the accuracy of the second-harmonic electrical technique applied to PMA samples (at least for PMA samples
in which the planar Hall effect is substantial) \cite{karimeddiny2021sagnac}.

\section*{Materials and Methods}
\subsection*{Sample fabrication}
The sample heterostructures are grown by DC-magnetron sputtering at a base pressure of less than 3$\times10^{-8}$ torr on high-resistivity, surface-passivated Si/SiO$_2$ substrates. Hall bars are patterned using photolithography and ion mill etching, then Ti/Pt contacts are deposited using photolithography, sputter deposition, and liftoff. The Co is deposited with a continuous thickness gradient (``wedge") across the 4-inch wafers and all devices measured have their current flow direction oriented along the thickness gradient.  The Hall-bar devices measured are 20 \textmu m $\times$ 80 \textmu m in size and the change in Co thickness is negligible on this scale i.e. the gradient over 80 \textmu m is orders of magnitude smaller than the RMS film roughness. The Ta underlayer is used to seed a smooth growth of subsequent films and the MgO/Ta forms a cap to minimize oxidation of the Co layer. 

\subsection*{Sagnac interferometer design}
\begin{figure*}[h]
    \centering
    \includegraphics[width=\linewidth]{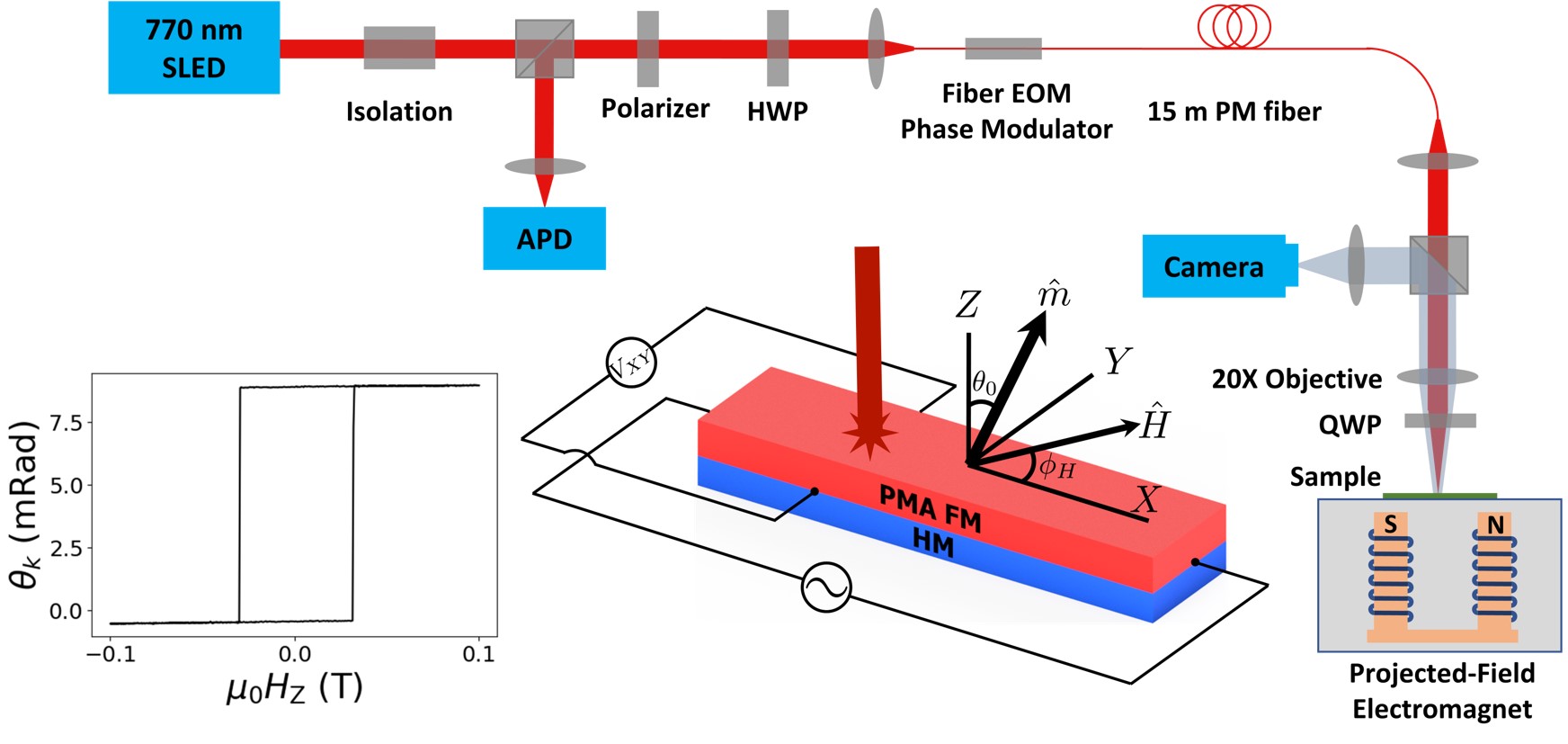}
    \caption{\textbf{Schematic of the Sagnac interferometer.} The left inset shows the Sagnac signal for out-of-plane magnetic-field-swept hysteresis of a Pt(4 nm)/Co(1.15 nm)/MgO device with $\mu_0 \meff \approx 0.42$ T; this is the same device for which we show data in Figs.~2. The middle inset depicts the device structure and coordinate definitions. 
        }
    \label{beamline}
\end{figure*}

Our Sagnac interferometer, modeled after those in refs.\ \cite{Xia2006,Fried2014}, is shown Fig.~\ref{beamline}. The beamline begins with a 770 nm superluminescent diode (SLED). The beam goes through a pair of Faraday isolators that provide $>65$ dB of backward isolation and prevent back-reflections into the diode that would cause intensity fluctuations and other source instabilities. Next, the beam goes through a beam splitter, polarizer, and half-wave plate (HWP) that prepare the beam polarization to be 45$^\circ$ with respect to the slow axis of a single mode polarization-maintaining (PM) fiber into which it is focused. The beam will henceforth be discussed as an equal combination of two separate beams of linearly-polarized light: one polarized along the slow axis and one polarized along fast axis of the PM fiber. A fiber electro-optic phase modulator (EOSPACE Inc.) applies time-dependent phase modulation to the beam traveling along the slow and fast axes with different amplitude modulation depths: $\phi_\parallel$ or $\phi_\perp$, respectively. The difference of these two amplitude modulation depths, $\phi_m = \phi_\parallel - \phi_\perp$ is controlled by a Lock-in oscillator voltage output (Zurich Instruments HF2LI). The beam then travels along 15 meters of PM fiber, whereupon it is collimated and focused by a long-working-distance objective through a quarter-wave plate (QWP) and onto a sample. The QWP is oriented such that one beam is converted to left-circularly-polarized light and the other is converted to right-circularly-polarized light. The beams then reflect off of a sample, exchanging the handedness of the beams and, if the sample is magnetic, imparting both the effects of circular dichroism and circular birefringence; the latter is equivalent to a Kerr rotation of linearly-polarized light. Upon reflection, the two beams (now exchanged) backpropagate and acquire a net phase difference of $\phi_m[\sin(\omega(t+\tau)-\sin(\omega(t)]$ at the EOM, where $\tau$ is the time it takes for the light to make the round trip back. The two beams interfere to produce homodyne intensity oscillations at the EOM frequency. The backpropagating beams are then routed by the beam splitter and focused into a broadband avalanche photodetector (APD). The APD's output voltage is measured by a lock-in amplifier that references the driving frequency of the EOM, $\omega$. To simplify the interpretation of the signal, the frequency $\omega$ is tuned such that $\omega = \pi/\tau$ \cite{Xia2006} [$2\pi(3.347 \text{ MHz})]$ for our apparatus). To maximize the Kerr rotation signal, the phase modulation depth $\phi_m$ is set by tuning the magnitude of AC voltage ($V_{pk} $ = 0.65 V) applied to the EOM so that $\phi_m = 0.92$ \cite{Fried2014}. With these simplifying calibrations, the Kerr rotation signal can be expressed as (see Supplementary Information section II for a full derivation)
\begin{align}\label{theta_k}
\begin{split}
    \theta_k 
    \approx \frac{1}{2}\arctan\left[0.543 \frac{V_\text{APD}^{\omega}}{V_\text{APD}^{2\omega}}\right],
\end{split}
\end{align}
where $V_\text{APD}^{\omega}(V_\text{APD}^{2\omega})$ is the APD voltage measured at the first- and second-harmonic of the EOM frequency. We quantify our Kerr rotation noise to be less than 5 \textmu Rad/$\sqrt{\text{Hz}}$ using a low power density on the sample (2 \textmu W/$\mu$m$^{2}$), comparable to the noise in ref.\ \cite{Fried2014} with the similar average power on the APD detector ($\sim$1 \textmu W). The low power ensures that the laser does not significantly heat the sample. More details can be found in the Supplementary Information sections II \& III.

\noindent\textbf{Acknowledgments:}
We acknowledge helpful discussions with Chenhao Jin, Kin Fai Mak, Yan S. Li, and Shengwei Jiang, and technical assistance from Vishakha Gupta, Rakshit Jain, Bozo Vareskic, and Reiley Dorrian.
We thank the the LASSP graduate student machine shop and its manager, Nathan I. Ellis, for advising on custom-machined parts made by S.K. and Y.K.L. 
\vspace{4mm}

 \noindent\textbf{Funding:} 
This work was funded by the National Science Foundation (DMR-1708499, DMR-2104268), the AFOSR/MURI project 2DMagic (FA9550-19-1-0390), and Task 2776.047 of ASCENT, one of six centers in JUMP, a Semiconductor Research Corporation program sponsored by DARPA.  Support from the NSF via the Cornell Center for Materials Research (CCMR) assisted in the construction of the Sagnac interferometer (DMR-1719875). Y.K.L. is supported by a Cornell Presidential Postdoctoral Fellowship and CCMR. T.M.J.C. is supported by the Singapore Agency for Science, Technology, and Research. The devices were fabricated using the shared facilities of the Cornell NanoScale Facility, a member of the National Nanotechnology Coordinated Infrastructure (supported by the NSF via grant NNCI-2025233) and the facilities of the CCMR.
\vspace{4mm}

\noindent\textbf{Author Contributions:}
S.K. and Y.K.L. devised the experiment and built the Sagnac apparatus. S.K., T.M.J.C., and Y.K.L. performed the measurements. T.M.J.C. fabricated the devices. S.K., T.M.J.C., and Y.K.L. performed the data analysis. O.S. performed quadrative MOKE calculation. D.C.R. provided oversight and advice. Y.K.L., D.C.R., T.M.J.C., O.S., and S.K. wrote the manuscript. All authors discussed the results and the content of the manuscript.
\vspace{4mm}

\noindent\textbf{Competing Interests:}
The authors declare no competing interests.
\vspace{4mm}
 
\noindent\textbf{Data and Materials Availability:}
All data needed to evaluate the conclusions in the paper are present in the paper and/or the Supplementary Materials.
\vspace{4mm}

\noindent\textbf{Correspondence:}
Correspondence and requests for materials should be addressed to D. C. Ralph and Y. K. Luo.



\newpage

\newpage

\begin{figure*}[h]
    \centering
    \includegraphics[width=\linewidth]{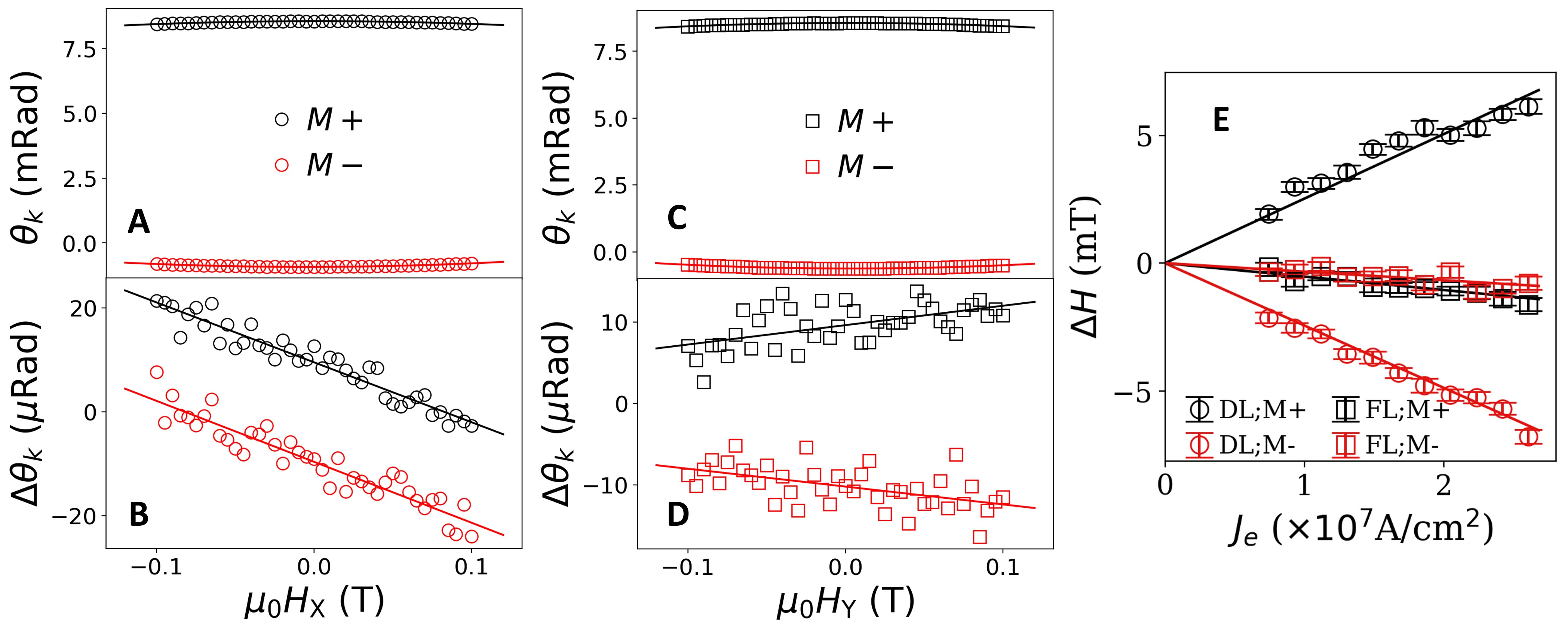}
    \caption{\textbf{Sagnac interferometry measurements of current-induced torque for a Pt(4 nm)/Co(1.15 nm)/MgO sample with perpendicular magnetic anisotropy.} (\textbf{A, B}) The Sagnac signals $\theta_k$ and $\Delta \theta_k$ for an in-plane magnatic field applied in the $X$ direction, for which $\Delta \theta_k$ provides a measurement of the damping-like torque.  (\textbf{C, D}) Corresponding signals for an in-plane magnetic field applied in the $Y$ direction, for which $\Delta \theta_k$ provides a measurement of the field-like torque. (\textbf{E}) Current-induced effective fields as a function of current density in the Pt layer, with linear fits to extract the spin-torque efficiencies. 
    } 
    \label{PMA Sagnac}
\end{figure*}
\newpage

\newpage
\begin{figure*}[h]
    \centering
    \includegraphics[width=\linewidth]{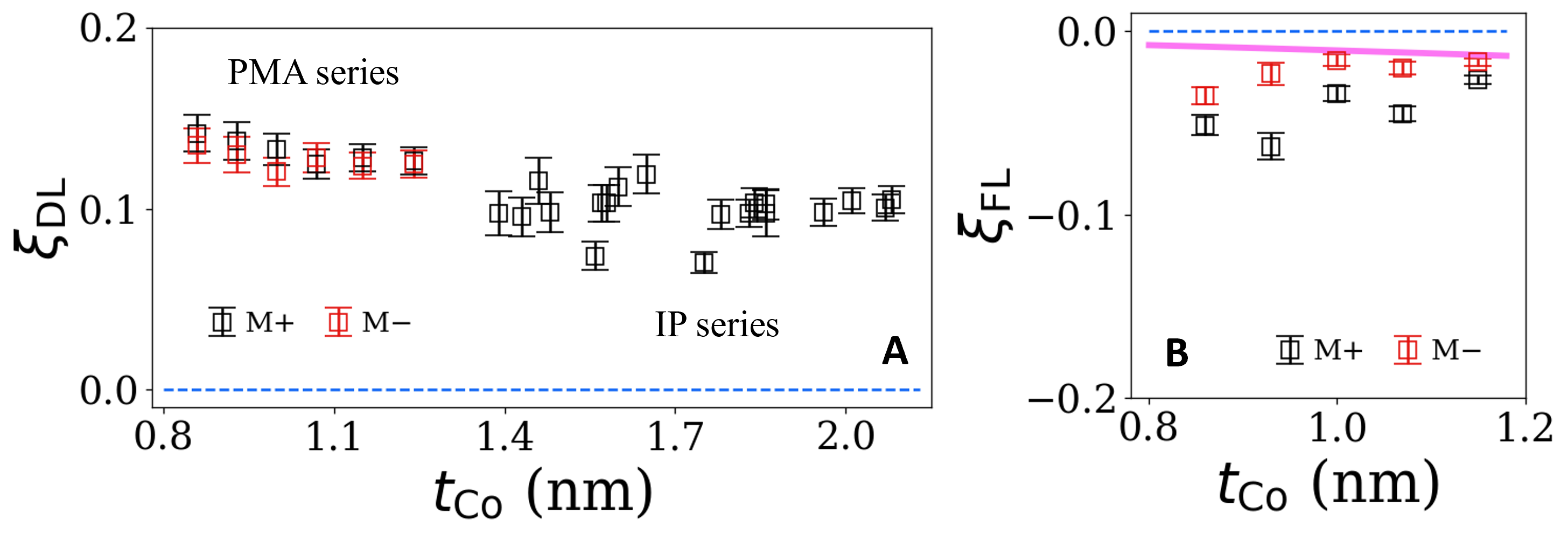}
    \caption{\textbf{Final results for the damping-like and field-like spin-orbit-torque efficiencies for the sample series Substrate/Ta(1.5)/Pt(4)/Co(0.85$-$2.1)/MgO(1.9)/Ta(2).} The numbers in parentheses are thicknesses in nanometers. The pink line in the (\textbf{B}) indicates the estimated Oersted torque based on current density. The larger error bars for the IP series compared to the PMA series in (\textbf{A}) are primarily a result of greater sample-to-sample scatter in the VSM measurements of $M_s t_\text{Co}$ rather than uncertainty in the Sagnac measurements. 
    }
    \label{torques}
\end{figure*}



\clearpage


\begin{center}
    \textbf{\LARGE{Supplementary Information}}\\
    \vspace{5 mm}
     \textbf{\LARGE{Sagnac interferometry for high-sensitivity optical measurements of spin-orbit torque}}\\

\vspace{5mm}    
\noindent\normalsize{Saba Karimeddiny\textsuperscript{1$\dagger$}, Thow Min Jerald Cham\textsuperscript{1$\dagger$}, Orion Smedley\textsuperscript{1}, Daniel C. Ralph\textsuperscript{1,2*},}\\
\noindent\normalsize{Yunqiu Kelly Luo\textsuperscript{1,2,3$\dagger$*}}\\
\vspace{5mm}
\small{\noindent\textit{\textsuperscript{1}Cornell University, Ithaca, NY 14850, USA}}\\
\textit{\textsuperscript{2}Kavli Institute at Cornell, Ithaca, NY 14853, USA}\\
\textit{\textsuperscript{3}Department of Physics and Astronomy, University of Southern California,} \\
\textit{Los Angeles, CA 90089, USA}\\
\textit{\textsuperscript{*}Corresponding authors. Email: dcr14@cornell.edu, kelly.y.luo@usc.edu}\\
\textit{\textsuperscript{$\dagger$}These authors contributed equally to this work.}
\end{center}

\maketitle
\newpage
\tableofcontents
\newpage

\renewcommand{\thesection}{\Roman{section}.} 
\renewcommand{\thesubsection}{\Alph{subsection}.}
\renewcommand{\thefigure}{S\arabic{figure}}
\setcounter{figure}{0}

\section{Details of the Sagnac Interferometer}
We begin this section by recommending the work by Fried et al., Rev.\ Sci.\ Instrum.\ {\bf 85}, 103707 (2014) \cite{Fried2014}. This paper served as the most helpful resource when building and debugging our interferometer, and many of the details in this section are inspired by the helpful level of detail in that work. We will show again here Fig.~\ref{beamline} from our main, but we will discuss some of the finer details of the apparatus.
\begin{figure}[h!]
    \centering
    \includegraphics[width=\linewidth]{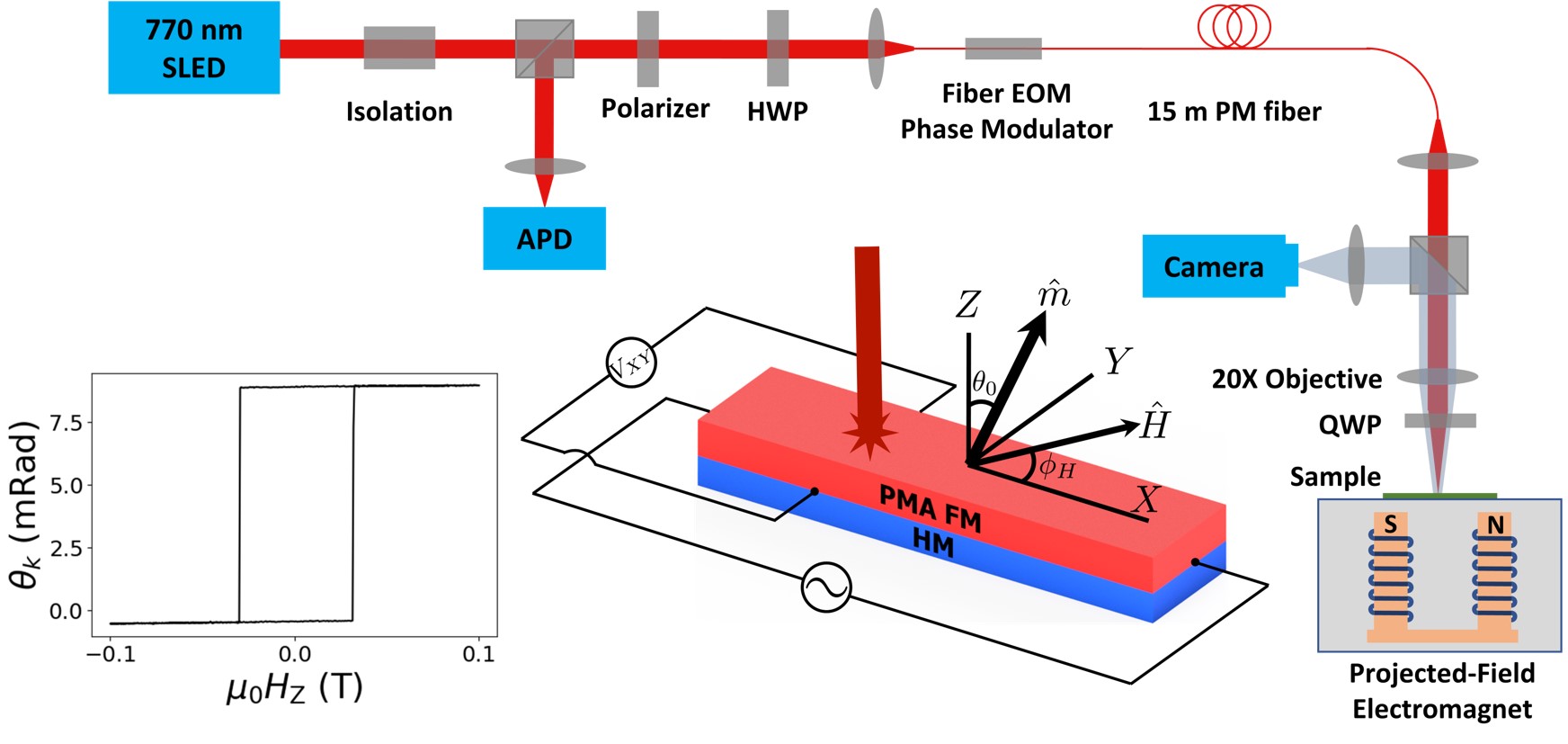}
    \caption{\textbf{Schematic of the Sagnac interferometer. } Main text Fig.~\ref{beamline} repeated here for ease of viewing.}
    \label{beam}
\end{figure}
The entire setup, including all of the optics, sample stage, and magnet are housed on a floating optical table and enclosed in a rigid polycarbonate box affixed with sound-proof foam to block air currents and external vibrations.

For the source, we use a 770 nm SLED, which has a broad ($\approx$ 15 nm) linewidth. In our original design we used an ultra-narrow-linewidth 780 nm diode, but we found that the broad-linewidth source reduced our noise by about a factor of two; this is because the small-linewidth source has a long beam coherence length. Therefore, the forward-going beam remained coherent with the reflected beam upon cycling through the apparatus, which gave them the opportunity to interfere and produce spurious interference signals not related to the Kerr rotation. The SLED diode and most of its pigtailed fiber are stored inside a closed styrofoam box within the polycarbonate box to further prevent temperature fluctuations and air currents.
We use two Faraday isolators that provide $>$ 65 dB of isolation to protect the diode from backreflections. Backreflections into the diode affect not only the longevity of the diode itself, but also can cause spurious intensity/spectral variations.

Next, the EOSpace fiber electro-optic phase modulator (EOM) is driven by a 50-MHz-bandwidth Zurich Instruments HF2LI lock-in amplifier and all of the signals (transport and optics) are detected on the same lock-in using its multiple demodulators. Our EOM is permanently pigtailed with a 5 m fiber and we append a 10 m fiber to it for a total length of 15 m. Both the EOM and the fiber are stored inside a closed styrofoam box (not the same box as the diode) within the polycarbonate box.

Upon exiting the fiber above the objective stage, the beam is collimated by a screw-on FC/APC lens adapter to a beam diameter of about 8 mm. We choose such a large beam diameter to maximize the filling of the back aperture of the objective lens and reduce our beam spot size on the sample. The beamsplitter after the collimating lens is retractable. It is illuminated with white light and inserted only to align the desired sample properly under the beam. The beam does not go through this beamsplitter during measurement. For the objective we choose to use a 20$\times$ near-IR ultra-long-working-distance objective lens to minimize the spot size, maximize the numerical aperture and field-of-view, and leave enough room for probes to make contact to the sample. The quarter-wave plate (QWP) is placed {\it after} the lens so that light is still linearly polarized while going through the lens. Most lenses have non-negligible Verdet constants so this is very important for reducing the spurious Faraday rotation incurred by the beam while it traverses the lens.

Our beam spot size on the sample is approximately 6 \textmu m and our power incident on the device is $<$ 70 \textmu W. We find that optical powers exceeding a few hundred \textmu W can begin to show local heating effects on the sample as indicated, e.g., as a change in the magnetic coercivity. To accommodate such a low-power beam, we detect the signal with a 50-MHz-bandwidth avalanche photodiode (APD) because it maintains a very low noise equivalent power (NEP) while sacrificing its saturation power, which we remain safely below. 

\section{Derivation of the Sagnac MOKE Signal}
\subsection{Measurement of the Kerr rotation angle $\theta_k$ in the absence of applied current}
\label{kerr section}
We will use the language of Jones matrices to derive Eq.~\ref{Sagnac Kerr} in the main text. This formalism allows us to calculate the behavior of two orthogonal modes of light at the same time.
First, we define some general Jones matrices: 
\begin{align}
\text{P}(\theta_p) &= \begin{pmatrix}
		 \cos^2\theta_p & \sin\theta_p\cos\theta_p \\
		 \sin\theta_p\cos\theta_p & \sin^2\theta_p
	          \end{pmatrix} \\
\text{WP}(\theta_{wp},\phi_{wp}) &= \begin{pmatrix}
		              \cos\frac{\phi_{wp}}{2} + i \sin\frac{\phi_{wp}}{2}\cos2\theta_{wp} & i \sin\frac{\phi_{wp}}{2}\sin2\theta_{wp} \\
		              i \sin\frac{\phi_{wp}}{2}\sin2\theta_{wp} & \cos\frac{\phi_{wp}}{2} - i \sin\frac{\phi_{wp}}{2}\cos2\theta_{wp}
	                       \end{pmatrix} \\
\text{EOM}(t) &= \begin{pmatrix}
		  e^{i \phi_\parallel \sin\omega t} & 0\\
		  0 & e^{i \phi_\perp \sin\omega t}
	          \end{pmatrix} \\
\text{S} &= \frac{1}{2}\begin{pmatrix}
		 \frac{e^{-i \delta_+}}{r_+} +  \frac{e^{-i \delta_-}}{r_-} &  i \left(\frac{e^{-i \delta_+}}{r_+} -  \frac{e^{-i \delta_-}}{r_-}\right)\\
		  -i \left(\frac{e^{-i \delta_+}}{r_+} -  \frac{e^{-i \delta_-}}{r_-}\right) & \frac{e^{-i \delta_+}}{r_+} +  \frac{e^{-i \delta_-}}{r_-}
	          \end{pmatrix}.	 
\label{eqn: optical matricies}
\end{align}
Here, our Jones vectors are in the basis of the laboratory: P($\theta_p$) is a polarizer oriented at an angle $\theta_p$. WP($\theta_{wp}$,$\phi_{wp}$) is a $\phi_{wp}$-wave plate oriented at an angle $\theta_{wp}$. EOM is the electro-optical phase modulator that applies a voltage-dependent phase ($\phi_\perp$ or $\phi_\parallel$ depending on whether the polarization of the incoming beam is along or perpendicular-to the optical axis of the EOM crystal) at a frequency of $\omega$. In the main text we say that the EOM only applies the phase to the beam traveling along the slow axis of the fiber; this is how the EOMs are designed, but our Jones matrix is more general to account for some phase shifts in the fast-axis beam, as well. Our final result is unchanged by this. S is the effect of the sample, which quite generally, has left- and right-circularly polarized light as its eigenvectors and applies an unequal phase ($\delta_+ \neq \delta_-$; ``circular birefringence") and an unequal Fresnel reflectance ($r_+ \neq r_-$; ``circular dichroism") to each of the two helicities of light. The effect of the sample reflectance exchanging the handedness of circularly polarized light is not captured by S, but will rather be accomplished by a complex conjugation later. 

At the start of the beam path for the interferometer, unpolarized light exits our laser and encounters a polarizer, P, oriented such the power lost through cross-polarization of the source beam is minimized (the source diode outputs partially-polarized light). We will assume without loss of generality that polarizer angle is 0$^\circ$ so the starting point for our Jones calculus is
\[
	v =\begin{pmatrix}
		1 \\
		0
		\end{pmatrix}.
\]
From our beam path we can simply apply the time-ordered Jones matrices of our optical components:
\begin{align}
 	\text{P}(0) \text{WP}(\pi/8,\pi) \text{EOM}(t + \tau) \text{WP}(\pi/4,\pi/2)\left[\text{S} \, \text{WP}(\pi/4,\pi/2) \text{EOM}(t) \text{WP}(\pi/8,\pi) v\right]^*.
\label{eqn: Jones}
\end{align}
In words, we begin with linearly-polarized light that is polarized at 0$^\circ$ ($v$). The beam goes through a half-wave plate that rotates the polarization of the beam to 45$^\circ$, which is equivalent to to two superimposed beams, one horizontally polarized and one vertically polarized.  Subsequently, the light goes through an EOM at time $t$, then through a quarter-wave plate, reflects from the sample, the LCP and RCP beams exchange due to the reflection (this is captured by the complex conjugation), goes through the quarter-wave plate again, through the EOM at a (now later) time $t + \tau$, and finally through the polarizer. We define $\tau$ as the time it takes for the beam to travel from the EOM to the sample and back. The result of the above matrix product is

\begin{align}
	\left(
 \frac{i e^{- i \delta_- +  \phi_\parallel \sin\omega t + \phi_\perp\sin\left[\omega(t+\tau)\right]}}{2\,r_-} + 
 \frac{i e^{- i \delta_+ + i \phi_\perp \sin\omega t + \phi_\parallel\sin\left[\omega(t+\tau)\right]}}{2\,r_+}\right)
	\begin{pmatrix}
	1 \\
	0
	\end{pmatrix}.
\label{eq: Ex}
\end{align}


In our experiment, we specifically tune the EOM frequency, $\omega$, such that $\tau = \pi/\omega$ \cite{Xia2006,Fried2014}; this results in the simplification:
\begin{align}
	\left(\frac{i e^{- i \delta_- + i \phi_m\sin\omega t}}{2\,r_-} + \frac{i e^{- i \delta_+ - i \phi_m\sin\omega t}}{2\,r_+}\right)
	\begin{pmatrix}
	1 \\
	0
	\end{pmatrix}
\label{eq: Ex}
\end{align}
where $\phi_m$ is the modulation depth $\phi_m := \phi_\parallel - \phi_\perp$. We detect the time-averaged intensity of light so we take half of the complex square of the above to get:
\begin{align}
	\frac{1}{8\,r_-^2} + \frac{1}{8\,r_+^2} + \frac{1}{8\,r_-r_+}\left(e^{i(\delta_+ - \delta_-)} e^{2i\phi_m\sin\omega t}+ e^{-i(\delta_+ - \delta_-)}e^{-2i\phi_m\sin\omega t}\right).
\end{align}
Next, we define 
$$  \theta_k := ( \delta_+ - \delta_-) /2 
\label{eq: def thetak}$$ Note $\theta_k$ is the angle  linearly polarized light would rotate after hitting the sample, as one can see by applying the sample matrix (Eq.~\ref{eqn: optical matricies}) to any linearly polarized Jones vector.

Applying this definition and the Jacobi-Anger expansion to the light-intensity expression, we get
\begin{align}
	\frac{1}{8\,r_-^2} + \frac{1}{8\,r_+^2} + \frac{1}{8\,r_-r_+}
	\left(e^{2i\theta_k} \sum_{n=-\infty}^{\infty} J_n(2\phi_m) e^{i n \omega t}+ e^{-2i\theta_k}\sum_{n=-\infty}^{\infty} J_n(2\phi_m) e^{-i n \omega t}\right).
\label{eq: Jacobi Anger}
\end{align}

To measure the first harmonic signal, we use a lock-in amplifier to isolate the component proportional to $\sin(\omega t)$ 

\begin{align}
\begin{split}
	I^\omega = &\frac{1}{T}\int_T dt \left[\frac{1}{8\,r_-^2} + \frac{1}{8\,r_+^2} + \frac{1}{8\,r_-r_+} \left(e^{2i\theta_k} \sum_{n=-\infty}^{\infty} J_n(2\phi_m) e^{i n \omega t}+ e^{-2i\theta_k}\sum_{n=-\infty}^{\infty} J_n(2\phi_m) e^{-i n \omega t}\right)\right]\sin\omega t \\
	=&\frac{1}{2 i T}\int_T dt \left[\frac{1}{8\,r_-r_+}
	\left(e^{2i\theta_k} \sum_{n=-\infty}^{\infty} J_n(2\phi_m) e^{i n \omega t}+ e^{-2i\theta_k}\sum_{n=-\infty}^{\infty} J_n(2\phi_m) e^{-i n \omega t}\right)\right]\left[e^{i\omega t} - e^{-i\omega t}\right] \\
	=&\frac{1}{2 i T}\int_T dt \left[\frac{1}{8\,r_-r_+}
	\left(e^{2i\theta_k} \sum_{n=-\infty}^{\infty} J_n(2\phi_m) \left(e^{i (n+1) \omega t} - e^{i (n-1) \omega t}\right) \right.\right. \\
	&\hspace{3.2cm}\left. \left. + e^{-2i\theta_k}\sum_{n=-\infty}^{\infty} J_n(2\phi_m) \left(e^{-i (n-1) \omega t} - e^{-i (n+1) \omega t}\right)\right)\right].
\end{split}
\end{align}
The only terms in the sums that will survive the integration are those for which the complex time-dependent exponentials are identically 1 (i.e. when $n+1 =0$ or $n-1=0$):
\begin{align}
	I^\omega &= \frac{1}{2 i T}\int_T dt \left[\frac{1}{8\,r_-r_+}
	\left[e^{2i\theta_k} \left( J_{-1}(2\phi_m) - J_1(2\phi_m)\right)+ e^{-2i\theta_k}\left(J_{1}(2\phi_m) - J_{-1}(2\phi_m)\right)\right]\right]\\
	&=\frac{1}{T}\int_T dt \frac{1}{8\,r_-r_+}
	\left[\sin2\theta_k \left( J_{-1}(2\phi_m) - J_1(2\phi_m)\right) \right] \\
	&=\frac{1}{8\,r_-r_+}
	\left[\sin2\theta_k \left( J_{-1}(2\phi_m) - J_1(2\phi_m)\right) \right] \\
	&=-\frac{\sin2\theta_k J_1(2\phi_m)}{4\,r_-r_+}.
\end{align}
Here we applied the $J_{-1} = -J_{1}$  property of the Bessel-$J$ functions. We can compute the second harmonic (the $\cos2\omega t$ component) using an analogous procedure

\begin{align}
	I^{2\omega} &=\frac{\cos2\theta_k J_2(2\phi_m)}{4\,r_-r_+}.
\end{align}
From these two expressions, we can solve for $\theta_k$ and also normalize out all of the dependencies on the Fresnel amplitude coefficients ($r_+$ and $r_-$) by simply taking the ratio of the two signals:
\begin{align}
	\theta_k = -\frac{1}{2}\arctan\left[\frac{J_2(2\phi_m) I^\omega}{J_1(2\phi_m) I^{2\omega}}\right].
\end{align}

For our measurements, we maximize the first harmonic signal, because it is proportional to the quantity we want to measure ($ \theta_k$). By tuning $\phi_m$ to maximize $J_1(2\phi_m)$, we get  $\phi_m = 0.92$ \cite{Fried2014} and $J_2(2\phi_m)/J_1(2\phi_m)\approx 0.543$. The above equation and aforementioned constant are exactly Eq.~\ref{Sagnac Kerr} in the main text.

\subsection{Measurement of changes in the Kerr angle $\Delta \theta_k$ due to current-induced magnetic deflections}
To derive a similar result with an AC applied current, we can begin at Eq.~(\ref{eq: Jacobi Anger}) with an added oscillation from a time-dependent $\theta_k$ that results from current-induced tilting of the magnetic moment at the current frequency $\omega_e$:
\begin{align}
    \frac{1}{8\,r_-r_+}
	\left(e^{2i(\theta_k + \Delta\theta_k\sin\omega_e t)} \sum_{n=-\infty}^{\infty} J_n(2\phi_m) e^{i n \omega t}+ e^{-2i(\theta_k + \Delta\theta_k\sin\omega_e t)}\sum_{n=-\infty}^{\infty} J_n(2\phi_m) e^{-i n \omega t}\right).
\end{align}
We can apply the Jacobi-Anger expansion again  
\begin{align}
	\frac{1}{8\,r_-r_+}
	\left(e^{2i\theta_k} \sum_{n,m=-\infty}^{\infty} J_n(2\phi_m)J_m(2\Delta\theta_k) e^{i (n \omega + m \omega_e) t}+ e^{-2i\theta_k}\sum_{n,m=-\infty}^{\infty} J_n(2\phi_m)J_m(2\Delta\theta_k) e^{-i (n \omega + m \omega_e) t}\right).
\end{align}
Now we  demodulate this signal at the sideband frequency $\omega \pm \omega_e$. We will only show the $\omega+\omega_e$ derivation for sign simplicity, but the result is identical for the upper and lower sidebands:
\begin{align}
\begin{split}
	I^{\omega+\omega_e} = &\frac{1}{T}\int_T dt \frac{1}{8\,r_-r_+}
	\left(e^{2i\theta_k} \sum_{n,m=-\infty}^{\infty} J_n(2\phi_m)J_m(2\Delta\theta_k) e^{i (n \omega + m \omega_e) t}+ \right.\\
	& \hspace{3.5cm}\left. e^{-2i\theta_k}\sum_{n,m=-\infty}^{\infty} J_n(2\phi_m)J_m(2\Delta\theta_k) e^{-i (n \omega + m \omega_e) t}\right)\times \cos\left(\omega t + \omega_e t\right) \\
	 = &\frac{1}{2T}\int_T dt \frac{1}{8\,r_-r_+}
	\left(e^{2i\theta_k} \sum_{n,m=-\infty}^{\infty} J_n(2\phi_m)J_m(2\Delta\theta_k) e^{i (n \omega + m \omega_e) t}+ \right.\\
	& \hspace{3.7cm}\left. e^{-2i\theta_k}\sum_{n,m=-\infty}^{\infty} J_n(2\phi_m)J_m(2\Delta\theta_k) e^{-i (n \omega + m \omega_e) t}\right)\times \left(e^{\omega t + \omega_e t}+e^{-\omega t - \omega_e t}\right)
\end{split}
\end{align}
Again, the only complex exponentials that will survive integration are the ones where the exponent is identically zero. This leaves us with:
\begin{align}
\begin{split}
	I^{\omega+\omega_e} = & \frac{1}{16\,r_-r_+}
	\left[e^{2i\theta_k}\left( J_{-1}(2\phi_m)J_{-1}(2\Delta\theta_k) + J_{1}(2\phi_m)J_{1}(2\Delta\theta_k)\right) + \right.\\
	& \hspace{2cm}\left. e^{-2i\theta_k}\left( J_{-1}(2\phi_m)J_{-1}(2\Delta\theta_k) + J_{1}(2\phi_m)J_{1}(2\Delta\theta_k)\right)\right]\\
	= &\frac{1}{4\,r_-r_+}
	  \cos2\theta_k J_{1}(2\phi_m)J_{1}(2\Delta\theta_k).
\end{split}
\end{align}
In our experiments $\Delta\theta_k$ is very small so we use that $J_1(x) \approx x/2$ for small $x$ 
\begin{align}
\begin{split}
	I^{\omega+\omega_e}
	= \frac{\cos2\theta_k J_{1}(2\phi_m)}{4\,r_-r_+}
	& \Delta\theta_k.
\end{split}
\end{align}
Finally, we take the ratio of this signal with the second harmonic (at $\omega$) derived earlier to reach a simple expression for the current-induced change in the Kerr signal
\begin{align}
\begin{split}
	\Delta\theta_k= \frac{J_2(2\phi_m)I^{\omega+\omega_e}}{J_1(2\phi_m)I^{2\omega}}.
\end{split}
\label{eq: delta theta_k}
\end{align}
All of the $\Delta\theta_k$ data presented are determined using this equation.

\section{Absence of Quadratic MOKE effects}
Quadratic MOKE (qMOKE) is a magneto-optic effect that is second-order in magnetization, specifically, the in plane moments.
This section justifies analytically and experimentally our main-text claim that qMOKE negligibly impacts our Sagnac signals, despite appearing in conventional polar-Kerr rotation measurements. In those measurements, linearly polarized light illuminates the sample at normal incidence. Upon reflection,  the polarization rotates by \cite{Fan2016}
\begin{align}
    \theta_{k,linear} = \kappa m_z + \beta_Q m_x m_y 
\label{eqn: qMOKE}
\end{align}
where $\kappa$ is still the MOKE coupling parameter,  $\beta_Q$ is the qMOKE coupling parameter and $\hat{m}$ is the magnetization unit vector. The components of $\hat{m}$ are defined in coordinates such that $z$ is still the film normal, but now $x$ lies along the plane of light polarization. 


In contrast, the next section derives the Sagnac signal to be
\begin{align}
 \theta_{k,Sag} =
\kappa m_z+ 2 \beta _Q   m_x m_y \sin \left(\kappa m_z\right) \left[0.71 
\cos \left(2 \kappa m_z\right)-0.62
\cos ^2\left(\kappa m_z\right)\right]+O\left(\beta _Q^2\right).
\label{eqn: qMOKE Sagnac}
\end{align}
The first term is equivalent to $\theta_k := ( \delta_+ - \delta_-)/2$ defined previously. The second term comes from qMOKE.  However, unlike $\theta_{k,linear}$,   the ratio of the qMOKE term to the polar MOKE term is only of order $\beta_Q \kappa/\kappa = \beta_Q$ (not $\beta_Q/\kappa$).  Since $\beta_Q$ is of order 10$^{-4}$ \cite{Fan2016}, the qMOKE contribution should be negligible compared to the polar MOKE contribution to the Sagnac Signal.

\subsection{Calculation of the qMOKE contribution}
The effect of qMOKE on the Sagnac signal can be derived by extending our Jones matrix calculation (Eq.~\ref{eqn: Jones}), with two changes: a new sample matrix and mirror operator:

\begin{align}
 	\text{P}(0) \text{WP}(-\pi/8,\pi) \text{EOM}(t + \tau) \text{WP}(-\pi/4,\pi/2)\text{mirror} \left[\text{S}_Q \, \text{WP}(\pi/4,\pi/2) \text{EOM}(t) \text{WP}(\pi/8,\pi) v\right].
\label{eq: JonesQuadratic}
\end{align}

First, we replace the sample reflection matrix S used previously, with a new sample reflection matrix S$_Q$ that includes the quadratic effects. We start with our original sample matrix (Eq.~\ref{eqn: optical matricies}), and for simplicity assume no circular dichroism ($ \xi := 1/r_+ = 1/r_- $) . We set the phase shifts to be equal ($\delta_+ = -\delta_- = \kappa m_z$) without loss of generality, because unequal phase shifts only cause a global phase, which is irrelevant and also can be absorbed into the prefactor $\xi$. Finally we add in the quadratic effect term, following Fan et al.\ \cite{Fan2016}. For the coordinate frame where the in-plane magnetization points along the x direction, the sample matrix including quadratic MOKE can be written

\begin{equation}
M_{k} = 
\xi  \left(\begin{array}{cc} \cos \left(\kappa m_z\right)+\frac{1}{2} \beta _Q \sin ^2\left(\theta _{\text{}}\right) & \sin \left(\kappa m_z\right) \\ -\sin \left(\kappa m_z\right) & \cos \left(\kappa m_z\right)-\frac{1}{2} \beta _Q \sin ^2\left(\theta _{\text{}}\right) \\\end{array}\right)
\end{equation}
where $\theta_\text{}$ is the polar angle of the magnetization.  Note that this matrix is identical to Fan et al.\ \cite{Fan2016} to lowest order in $\kappa$ and $\beta_Q$.  
Next, because the magnetic moment may point in other directions besides the x-z-plane we change the basis of $M_K$ by a rotation  about the z axis \cite{Fan2016}. To do this, we apply the rotation matrix $R$, defined as 
\begin{equation}
R(\phi_\text{}) =\left[\begin{array}{cc}
\cos \phi_\text{} & -\sin \phi_\text{} \\
\sin \phi_\text{} & \cos \phi_\text{}
\end{array}\right],
\end{equation}   
where $\phi_\text{}$ is the azimuthal angle of the magnetization, to yield the general sample reflection matrix (in the linearly-polarized laboratory basis)

\begin{align}
\begin{split}
S_Q &= R(\phi_\text{}) \cdot M_K \cdot R(-\phi_\text{}) \\
&=
\xi \times \left(\begin{array}{cc} \cos \left(\kappa m_z\right)+\frac{1}{2} \beta _Q \sin ^2\left(\theta _{\text{}}\right) \cos \left(2 \phi _{\text{}}\right) & \sin \left(\kappa m_z\right)+\beta _Q \sin ^2\left(\theta _{\text{}}\right) \sin \left(\phi _{\text{}}\right) \cos \left(\phi _{\text{}}\right) \\ \beta _Q \sin ^2\left(\theta _{\text{}}\right) \sin \left(\phi _{\text{}}\right) \cos \left(\phi _{\text{}}\right)-\sin \left(\kappa m_z\right) & \cos \left(\kappa m_z\right)-\frac{1}{2} \beta _Q \sin ^2\left(\theta _{\text{}}\right) \cos \left(2 \phi _{\text{}}\right) \\\end{array}\right)
\end{split}
\end{align}
Note that for ferromagnets with PMA or easy-plane anistropy where the in-plane magnetization follows the in-plane applied field such as in our case: $\phi_\text{}$ = $\phi_\text{H}+\phi_\text{frame}$.  $\phi_\text{frame}$ is added to account for the arbitrary rotation of the reference frame of light polarization  relative to the current direction due to  the fiber. The above equation can also be expressed equivalently in terms of the Cartesian components of the magnetization unit vector: 
\begin{align}
    S_Q =
\xi \times \left(\begin{array}{cc} \cos \left(\kappa m_z\right) + \frac{1}{2} \beta _Q \left(m_x^2-m_y^2\right) & m_x m_y \beta _Q+\sin \left(\kappa m_z\right) \\ m_x m_y \beta _Q-\sin \left(\kappa m_z\right) & \cos \left(\kappa m_z\right) - \frac{1}{2} \beta _Q \left(m_x^2-m_y^2\right) \\\end{array}\right).
\end{align}

The second change between Eq.~(\ref{eqn: Jones}) and Eq.~(\ref{eq: JonesQuadratic}) is that the mirror operator replaces the complex conjugation. qMOKE requires this generalization because the complex conjugate only mirrors circularly polarized light, not linearly polarized light. qMOKE produces linear light components, even when illuminated with circular light, so it is necessary to mirror those components upon reflection as well. The mirror operator is 
\begin{align}
    \text{mirror} = \left(\begin{array}{cc}
1 & 0 \\
0 & -1
\end{array}\right).
\end{align}
which flips the sign of the y-component of the electric field. This choice of mirroring axis is physically irrelevant; it only causes a global phase shift and changes the angles of the optical elements through which the light back-propagates. We chose to mirror over the x axis, so that the angles of the optical elements simply pick up a minus sign.

The above two changes to the Jones matrices result in the following Jones vector at the detector:
\begin{equation}
-i \xi \left(\cos \left(\phi _m \sin (t w)+\kappa m_z\right)+\frac{1}{2} \beta _Q \sin ^2\left(\theta _{\text{}}\right) \sin \left(2 \phi _{\text{}}\right)\right)
	\begin{pmatrix}
	1 \\
	0
	\end{pmatrix}.
\end{equation}
The first term here matches the previously calculated version  (Eq.~(\ref{eq: Ex})), within the assumptions we made. The second term represents the quadratic effects, which have no dependence on the time-dependent phase modulation from the electro-optic modulator. If we were able to make a direct measurement of the modulated part of the electric field, qMOKE would give no contribution.  However, we measure the intensity, so the qMOKE signal contributes to the modulated intensity upon multiplying with the other term while taking the complex square.  Nevertheless, this results in a much smaller contribution to the Sagnac signal from qMOKE than might be anticipated intuitively.

Now that we have the Jones vector incident on the detector, we repeat the mathematics of section \ref{kerr section} to derive the new Sagnac signal:
\begin{equation}
\theta_{k,Sag}=
\frac{1}{4} \tan ^{-1}\left(\frac{2 \phi _m J_2\left(2 \phi _m\right) \sin \left(\kappa m_z\right) \left(\beta _Q J_1\left(\phi _m\right) \sin ^2\left(\theta _{\text{}}\right) \sin \left(2 \phi _{\text{}}\right)+J_1\left(2 \phi _m\right) \cos \left(\kappa m_z\right)\right)}{\left| \phi _m\right| J_1\left(2 \phi _m\right) \left(2 \beta _Q J_2\left(\phi _m\right) \sin ^2\left(\theta _{\text{}}\right) \sin \left(2 \phi _{\text{}}\right) \cos \left(\kappa m_z\right)+J_2\left(2 \phi _m\right) \cos \left(2 \kappa m_z\right)\right)}\right).
\label{eq: thetaSagnac}
\end{equation}
In the case of no quadratic effects ($\beta_Q=0$), this expression reduces to the first order Kerr rotation:
\begin{equation}
    \theta_{k,Sag} = m_z \kappa.
    \label{eq: thetaSagnac NoBeta}
\end{equation}
We can also expand the full expression for $\theta_{k,Sag}$ to first order in $\beta_Q$:
\begin{equation}
    \theta_{k,Sag}=
    \kappa m_z+\beta _Q \sin ^2\left(\theta _{\text{}}\right) \sin \left(2 \phi _{\text{}}\right) \sin \left(\kappa m_z\right) \left(\frac{J_1\left(\phi _m\right) \cos \left(2 \kappa m_z\right)}{J_1\left(2 \phi _m\right)}-\frac{2 J_2\left(\phi _m\right) \cos ^2\left(\kappa m_z\right)}{J_2\left(2 \phi _m\right)}\right)+O\left(\beta _Q^2\right)
\end{equation}
and, upon substituting in the modulation depth $\phi_{m} = 0.92$ used for the Sagnac measurement,
\begin{align}
 \theta_{k,Sag} =
\kappa m_z+\beta _Q \sin ^2\left(\theta _{\text{}}\right) \sin \left(2 \phi _{\text{}}\right) \sin \left(\kappa m_z\right) \left(0.71 
\cos \left(2 \kappa m_z\right)-0.62
\cos ^2\left(\kappa m_z\right)\right)+O\left(\beta _Q^2\right).
\label{eq: qMOKE taylor0.92}
\end{align}

From this we conclude that a qMOKE signal would have a dependence on the angle of the in-plane magnetization $\propto m_x m_y \propto \sin(2\phi_\text{})\propto \sin(2\phi_\text{H})$, and we have reached the expression of Eq.~(\ref{eqn: qMOKE Sagnac}).

\subsection{Experimental limit on the qMOKE contribution}
\begin{figure}[h!]
    \centering
    \includegraphics[width=\linewidth]{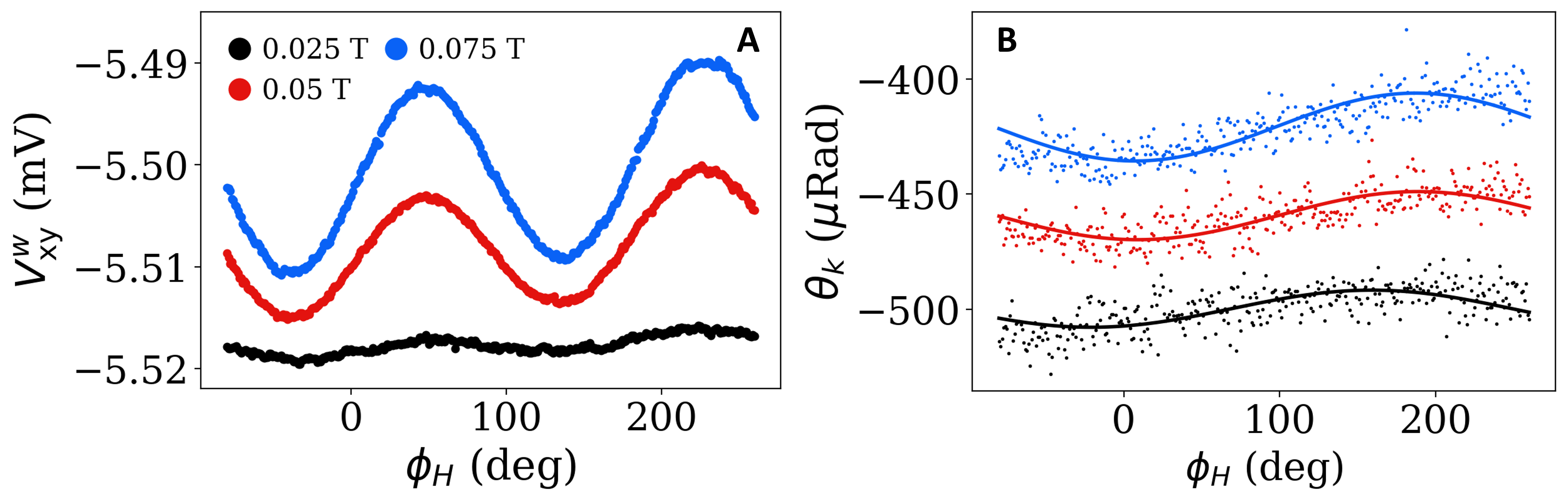}
    \caption{\textbf {Comparison between electrical Hall signal and Sagnac signal for in-plane field angle $\phi_H$ sweep.} ({\textbf A}) The measured first-harmonic electrical Hall signal, $V^\omega_{XY}$ due to a combination of the anomalous and planar Hall effects a{(\textbf B)} the Kerr rotation signal measured by the Sagnac interferometer, $\theta_k$, vs.\ the angle of applied magnetic field within the plane of the magnet. Three different strengths of applied magnetic field are shown and they are artificially vertically offset in {(\textbf B)} for clarity. The overlayed lines are best-fits to Eq.~\ref{PMAAng}.
  The data are collected for the Pt(4 nm)/Co(1.15 nm) sample with $\mu_0 \meff \approx 0.42$ T; this is the same device for which measurements are highlighted in the main text Fig.~\ref{PMA Sagnac}.}
    \label{quadmoke}
\end{figure}

To test experimentally for any contribution of qMOKE to the Sagnac signal, we perform a measurement analogous to a calibration of the electrical planar Hall effect -- we  directly measure the change in the Sagnac signal as we apply an in-plane magnetic field to tilt the PMA magnet partially in-plane and then rotate the field angle $\phi_H$. Supplementary Fig.~\ref{quadmoke} shows the result of these measurements for (a) the electrical Hall signal and (b) the Sagnac signal, each for 3 different strengths of applied magnetic field. The data for this figure were collected on the same Pt(4 nm)/Co(1.15 nm) sample discussed in the main text (i.e, Figs.~\ref{beamline}-\ref{PMA Sagnac}), which had $\mu_0 \meff \approx 0.42$ T, as calculated by the parabolic fits in the main text.

As explained above (Eq.~(\ref{eqn: qMOKE Sagnac})), if there were any measurable contribution from qMOKE, we should expect a signal $\propto \sin(2\phi_\text{H})$, i.e., with a $\pi$ periodicity.  
Such a $\pi$-periodic signal is clearly visible in the electrical planar Hall measurement.  However the $\phi_\text{H}$ dependence of the Sagnal signal is much weaker, and it is 2$\pi$-periodic, not $\pi$-periodic.  Therefore, no contribution from qMOKE is measurable.

Based on the calculation above (Eq.~(\ref{eqn: qMOKE Sagnac}) or (\ref{eq: qMOKE taylor0.92})), with the parameters $\kappa = 4.9 \times 10^{-3}$  (from the hysteresis curve in Fig.~\ref{beamline} of the main text), $\beta_Q = 1.1\times 10^{-4}$ (based on data for Pt/Py from X.\ Fan et al.\ \cite{Fan2016}), and $\theta_\text{} = \arcsin(H/\meff)$, the expected amplitude of the $\sin(2\phi_\text{H})$ signal for the $|\mu_0 H|$ = 75 mT scan  is approximately 1.4 nano-radians.  This is indeed orders of magnitude less than the experimental noise in Fig.~\ref{quadmoke}(b), so in agreement with the experiment we should not expect any visible qMOKE contribution.

\begin{figure}[h!]
    \centering
    \includegraphics[width=0.5\linewidth]{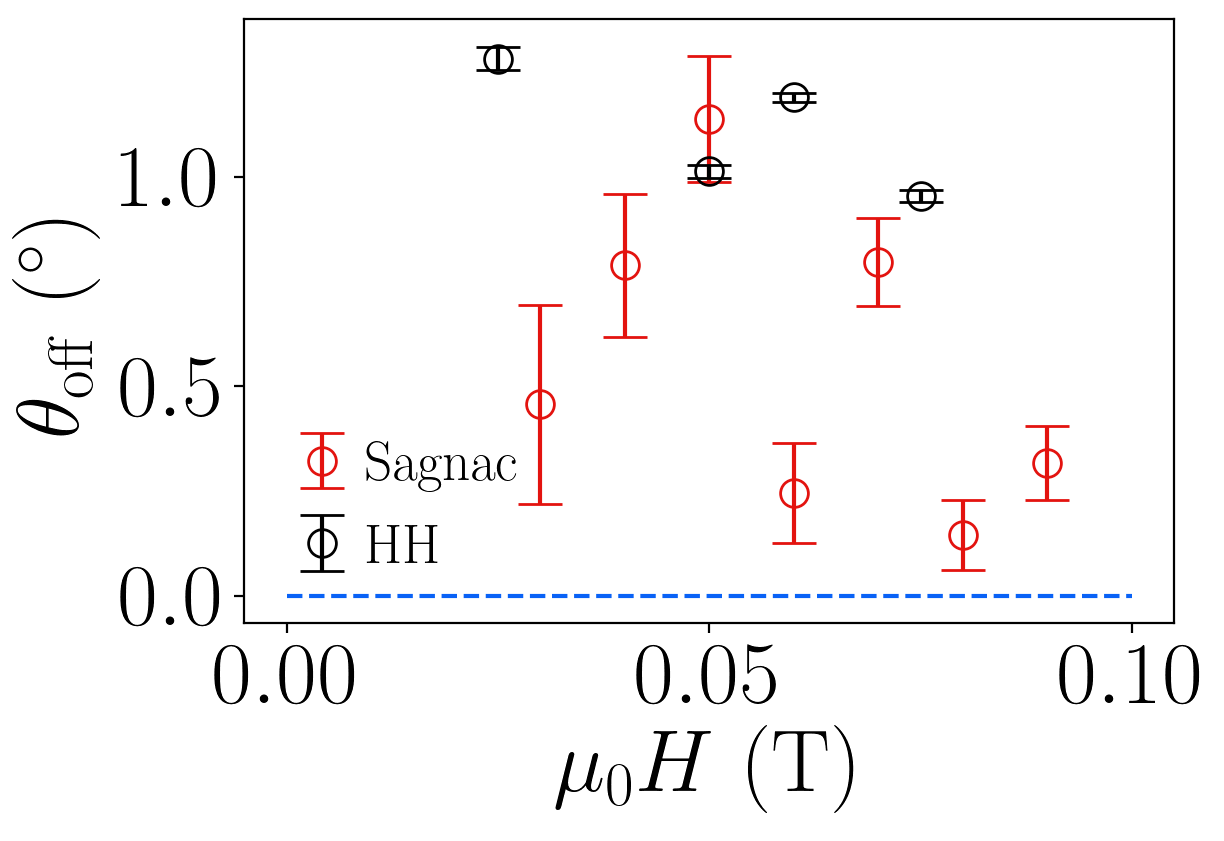}
    \caption{{\textbf{Field misalignment angle calibration.}} The measured field misalignment angle, $\theta_\text{off}$, versus the strength of applied magnetic field applied nominally in the plane of the device. $\theta_\text{off}=0^\circ$ means that the magnetic field is perfectly aligned in the device plane.}
    \label{strayField}
\end{figure}

The $\pi$-periodic signal that is visible in Fig.~\ref{quadmoke}(b) can be understood instead as due to a small misalignment of the applied magnetic field from the plane of the sample. The first-harmonic Hall voltage signal as a function for small tilt angles has the form \cite{Hayashi2014,karimeddiny2021sagnac}
\begin{align}
    \frac{V^{\omega}_{XY}}{\Delta I} =  &R_\text{AHE}\cos\left({\frac{H}{\meff}}\right) \nonumber\\  
    +&  R_\text{PHE}\sin^2\left({\frac{H}{\meff}}\right)\sin\phi_H\cos\phi_H \nonumber \\
   +&  R_\text{AHE}\frac{H^2\sin\theta_\text{off}}{(\meff)^2}\sin\left({\frac{H}{\meff}}\right)\cos(\phi_H - \phi_\text{off}). 
\label{PMAAng_R}
\end{align}
In analogy with Hall measurements, for a field rotation axis misaligned by an angle $\theta_\text{off}$ relative to the sample normal direction, the polar-MOKE Sagnac signal should have the dependence for small tilts \cite{karimeddiny2021sagnac}
\begin{equation}
    \theta_k \approx \kappa \cos\left({\frac{H}{\meff}}\right)   
   + \kappa \frac{H^2\sin\theta_\text{off}}{(\meff)^2}\sin\left({\frac{H}{\meff}}\right)\cos(\phi_H - \phi_\text{off}).
\label{PMAAng}
\end{equation}
Fits of these curves for both the Hall and Sagnac measurements are shown in Supplementary Fig.~\ref{quadmoke}.
Both sets of data indicate a field/sample tilt of  $\theta_\text{off} \sim 1^\circ$ (Supplementary Fig.~\ref{strayField}). This is most likely caused by a slight misalignment of the projected-field magnet (GMW 5201) center. 

\newpage
\section{Expression of Kerr rotation $\theta_K$ and current induced changes in the Kerr angle $\Delta\theta_K$.}
In this section, we will derive the field dependence of the Kerr rotation and current induced changes in the Kerr angle, coupled to just the out-of-plane component of the magnetic moment. To do this, we follow a procedure similar to those used in refs.\ \cite{Hayashi2014,karimeddiny2021sagnac} for deducing the equilibrium positions and current-induced modulation amplitudes of the magnetization.

We begin the derivation by writing the equilibrium magnetic energy divided by the total magnetic moment in the absence of any applied current
\begin{align}
    \frac{F_\text{eq}(\theta,\phi)}{M_s} &= -\mu_0\mathbf{m} \cdot \mathbf{H} - \frac{\mu_0 M_\text{eff}}{2} (\mathbf{m}\cdot \hat{Z})^2 \nonumber\\
    &= - \mu_0H \sin\theta\sin\theta_H\cos(\phi-\phi_H) - \frac{\mu_0}{2}\cos\theta(2H\cos\theta_H+M_\text{eff}\cos\theta).
    \label{F}
\end{align}
Here $F_\text{eq}$ is the equilibrium free energy, $M_s$ is the saturation magnetization, $\mathbf{m}$ is the normalized vector magnetic moment, $\mathbf{H}$ is the vector external magnetic field, and $\mu_0\meff = \mu_0 M_s - 2K_\perp/M_s$
is the effective magnetization. PMA is indicated by a \textit{positive} $\meff$. The angles in the second line denote the direction of external applied magnetic field when subscripted with an $H$ and refer to the the  direction of the magnetic moment when they lack a subscript.  Minimization of this free energy yields the equilibrium magnetic orientation $\theta_0$,$\phi_0$. As we apply an AC current the SOTs produced will act as effective fields that reorient the magnetic moment. This is a ``slow" process ($\dot{\mathbf{m}} \ll \gamma |H|$) so it may be described as an effective modification of equilibrium free energy (Eq.~(\ref{F})). With the perturbation from a general, current-induced effective magnetic field, $\Delta\mathbf{H}$ (assumed small compared to $H$), the free energy becomes
\begin{align}
\begin{split}
    \frac{F(\theta,\phi)}{M_s} =&  \frac{F_\text{eq}(\theta,\phi)}{M_s}   - \mu_0 \mathbf{m}\cdot\Delta \mathbf{H}\\
    \approx & \frac{F_\text{eq}(\theta_0,\phi_0)}{M_s} + \frac{1}{2M_s}\dfrac{\partial^2 F_\text{eq}}{\partial \theta^2}\Bigr|_{\theta_0,\phi_0} (\Delta \theta)^2  + \frac{1}{2M_s}\dfrac{\partial^2 F_\text{eq}}{\partial \phi^2}\Bigr|_{\theta_0,\phi_0} (\Delta \phi)^2 + \frac{1}{M_s}\dfrac{\partial^2 F_\text{eq}}{\partial \theta \partial \phi}\Bigr|_{\theta_0,\phi_0} \Delta \theta \Delta \phi \\
    &- \mu_0( \sin\theta \cos\phi \Delta H_X + \sin\theta \sin\phi \Delta H_Y + \cos\theta \Delta H_Z).
\end{split}
\end{align}
(The first derivatives of $F_\text{eq}$ are zero when evaluated at the equilibrium orientation.)  We have included the cross second derivative in this expression, but when evaluated it gives zero. 

The new magnetic orientation in the presence of the current-induced magnetic field can then be calculated as a minimization problem
\begin{align}
\dfrac{\partial F}{\partial \theta} = \dfrac{\partial F}{\partial \phi} = 0.
\label{MinEq}
\end{align}
Here we consider the case where the external field is in-plane ($\theta_H$ = $\frac{\pi}{2}$) and assume negligible within-plane anisotropy so that the in-plane projection of the equilibrium magnetic moment is aligned with the external field i.e. $\phi_0 = \phi_H$.

The solutions of Eq.~(\ref{MinEq}) to first order in the current-induced field yield
\begin{align}
    \Delta\theta &= \frac{\cos\theta_0(\Delta H_X \cos\phi_H + \Delta H_Y \sin\phi_H) - \Delta H_Z\sin\theta_0}{M_\text{eff}\cos2\theta_0 + H\sin\theta_0} 
    \label{DeltaTheta} \\
    \Delta\phi &= \frac{-\Delta H_X\sin\phi_H + \Delta H_Y \cos\phi_H}{H}.
    \label{DeltaPhi}
\end{align}
The solution for the equilibrium polar angle of the magnetization is
\begin{align}
 \dfrac{\partial F_{eq}}{\partial \theta}\Bigr|_{\theta_0,\phi_0} &= -\mu_0H \cos\theta_0 - \mu_0M_\text{eff}\sin\theta_0\cos\theta_0 = 0\\
\theta_0 &= \begin{cases} \arcsin (\frac{H}{M_\text{eff}}) & H < M_\text{eff}\\
\frac{\pi}{2} & H \geq M_\text{eff}.
\end{cases}
\label{eq:arcsin}
\end{align}
To get from the above equations to the full expected Sagnac MOKE signal (Eqs.\ (4), (5) and (9) in the main text) we begin with the expression for the polar MOKE signal in the linear regime, which is only sensitive to the out-of-plane component of the magnetization m$_z$
\begin{align}
    \theta_k &= \kappa m_Z.
\end{align}
We let $\theta$ consist of an equilibrium contribution due to the external magnetic field and a time-dependent contribution due to the AC-current-induced spin-orbit fields: $\theta \to \theta_0 + \Delta\theta$ with $\Delta\theta \ll 1$ and then Taylor expand
\begin{align}
    \theta_k + \Delta\theta_k \approx \kappa \left(\cos\theta_0 - \Delta\theta\sin\theta_0\right).
    \label{thetaKExpand}
\end{align}
For the tilting measurements on samples with perpendicular magnetic anisotropy, $\sin\theta_0 = H/M_\text{eff}$  (Eq.~\ref{eq:arcsin}) and $\phi_0 = \phi_H$. Using the expression for $\Delta\theta$ derived previously (Eq.~\ref{DeltaTheta}), and separating the equilibrium and current-induced signals in Eq.~(\ref{thetaKExpand}), in the regime of weak applied fields $H \ll M_\text{eff}$ we get for small tilts about the $\pm$ m$_z$ directions
\begin{align}
    \theta_k = &\pm\kappa\left(1 - \frac{H^2}{2M_\text{eff}^2}\right)\\
    \Delta\theta_k = & \mp\kappa\left(\Delta H_X \cos\phi_H +\Delta H_Y \sin\phi_H \right)\frac{H}{M_\text{eff}^2}.
\end{align}
For in-plane field azimuth sweep measurements on samples with in-plane magnetic anisotropy, $\theta_0$ = $\frac{\pi}{2}$ and $\phi_0 = \phi_H$. Substituting in the expression for $\Delta\theta$ from Eq.~(\ref{DeltaTheta}), and considering the current-induced term in Eq.~(\ref{thetaKExpand}), we get
\begin{align}
    \Delta\theta_K &= \frac{\kappa\Delta H_Z}{H-M_\text{eff}} \nonumber \\
    &= \frac{\kappa\Delta H_{DL}\cos\phi}{H-M_\text{eff}}.
    \label{DeltaThetaK}
\end{align}

\section{Comparison of a Sagnac MOKE measurement and a conventional polar-MOKE measurement for a PMA thin film}
Here we present a comparison between the conventional polar MOKE method and Sagnac MOKE interferometry for Ta/Co$_{40}$Fe$_{40}$B$_{20}$ test samples with perpendicular magnetic anisotropy. In Supplementary Fig.~\ref{MokeSagnacComparison}(a), out-of-plane magnetic hysteresis is measured using a conventional polar MOKE setup for a Ta(4 nm)/Co$_{40}$Fe$_{40}$B$_{20}$(0.65 nm) sample. In comparison, Supplementary Fig.~\ref{MokeSagnacComparison}(b) shows Sagnac MOKE interferometry readout on a similar CoFeB sample from the same wedge wafer with a slightly different thickness Ta(4 nm)/Co$_{40}$Fe$_{40}$B$_{20}$(0.85 nm). The difference in the coercivity field between Supplementary Fig.~\ref{MokeSagnacComparison}(a) and \ref{MokeSagnacComparison}(b) is due to this small difference in film thickness. One can visually observe that the signal-to-noise of the Sagnac MOKE interferometry is a significant improvement compared to conventional polar MOKE. The linear background in polar MOKE readout (Supplementary Fig.~\ref{MokeSagnacComparison}(a)) comes from the Faraday effect in the objective lens. Sagnac interferometry is insensitive to this effect in the case that the quarter-waveplate  is positioned between sample and the objective \cite{Fried2014}. Furthermore in this data, the Sagnac took a single scan in $\sim$ 1 minute with a lock-in amplifier time constant of 10 milliseconds, while the polar-MOKE took the average over 10 scans in a total of  $\sim$ 10 - 20 minutes, with a lock-in amplifier time constant of 500 milliseconds.  

Our conventional polar-MOKE setup mimics the setup of \cite{Kato2004} with the only difference of using a Helium-neon laser as the light source. A linearly-polarized beam is incident normal to the sample through an objective lens with a numerical aperture of 0.4, focusing the beam to a circular spot with a full width at half maximum of $\sim$ 1 \textmu m. The rotation of the reflected beam polarization is detected by a balanced photodiode bridge with a noise equivalent power of 1.1 pW/$\sqrt{\text{Hz}}$, which gives a readout noise of $\sim$ 400 \textmu Rad/$\sqrt{\text{Hz}}$ for the conventional MOKE setup (Supplementary Fig.~\ref{MokeSagnacComparison}(a)). In contrast, our Sagnac MOKE readout noise (Supplementary Fig.~\ref{MokeSagnacComparison}(b)) is less than 5 \textmu Rad/$\sqrt{\text{Hz}}$.
As we noted in the main text, while conventional MOKE can achieve comparable resolution with external modulation of magnetic field, electric field, or current \cite{Kato2004,Lee2017}, these methods are not applicable for measuring hysteresis curves of ferromagnets.

\begin{figure}[h!]
    \centering
    \includegraphics[width=\linewidth]{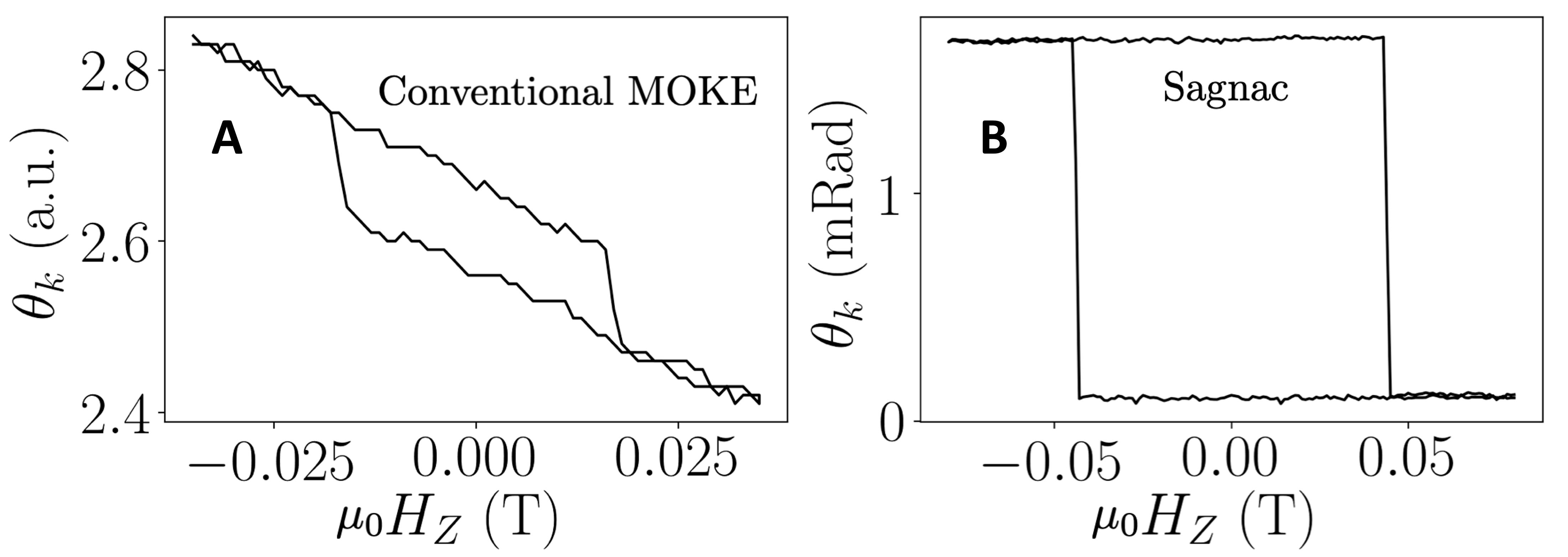}
    \caption{\textbf {Comparison between conventional MOKE and Sagnac readouts.} {(\textbf A)} Conventional polar MOKE readout v.s. {(\textbf B)} Sagnac MOKE interferometry readout on a CoFeB film with perpendicular magnetic anisotropy.}
    \label{MokeSagnacComparison}
\end{figure}
\section{P\MakeLowercase{t} Sample Details}
\subsection{Wedge Thickness}
All samples measured are grown by DC-magnetron sputtering onto a high-resistivity, surface-passivated Si/SiO$_2$ wafer. The stacks are Si/SiO$_2$/Ta(1.5)/Pt(4)/Co($t_\text{Co}$)/MgO(1.9)/Ta(2) where all of the numbers in parentheses are layer thicknesses in nanometers. The bottom Ta is used as a seed layer to promote smooth growth of the films, and the top MgO/Ta stack is used to cap the Co and minimize oxidation of the Co Layer. Both the bottom and capping Ta layers are sufficiently resistive that they carry negligible current density compared to the Pt and Co layers. By strategically stopping the wafer rotation during sputter deposition, we grow the Co layer with a thickness-gradient ``wedge". The wedge's thickness gradient is along the direction of current flow (X-axis) for all devices measured. The Co thickness as a function of device distance from the wafer flat is shown in Supplementary Fig.~\ref{tco}, for both PMA series and IP series shown in the main text. This calibration is performed using atomic-force microscopy measurements at different points on a test wafer, followed by a polynomial fit and interpolation to get the thickness variation across the full wafer.

\begin{figure}[h!]
    \centering
    \includegraphics[width=1\linewidth]{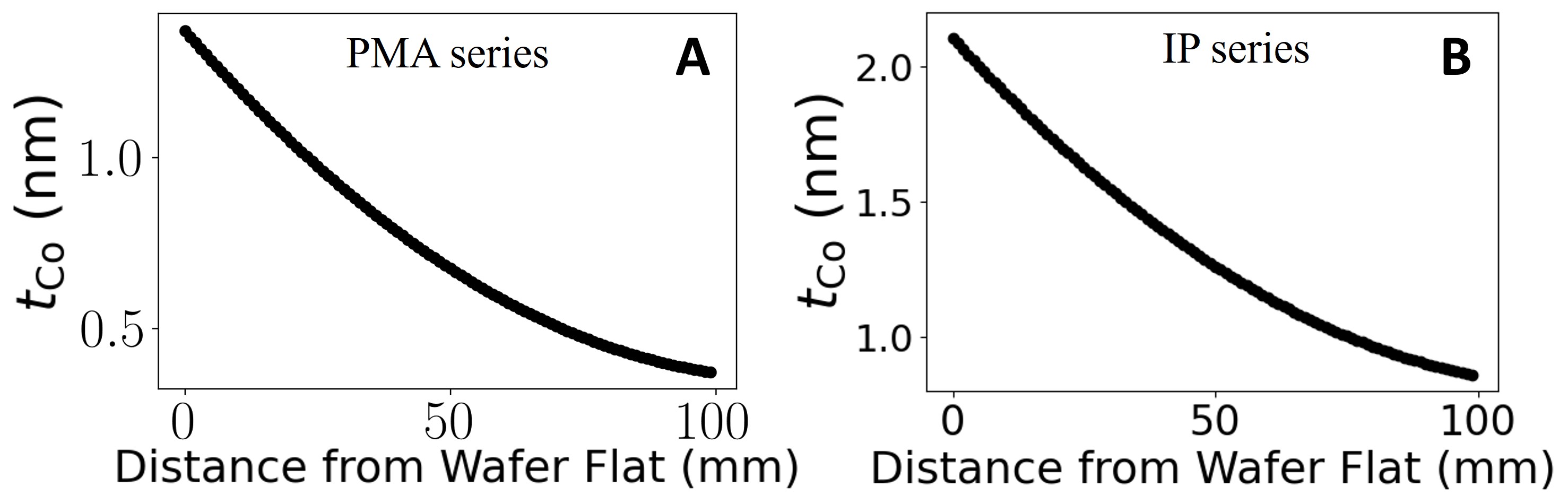}
    \caption{\textbf{Co thin wedge film calibration.} The thickness of the Co ``wedge" film as a function of the distance from the 4-inch wafer flat for \textbf{(A)} PMA series and \textbf{(B)} IP series.}
    \label{tco}
\end{figure}

\subsection{Film Conductivity}

We characterize the electrical conductances of our films by measuring the four-point resistance on many devices across the Co-wedge wafer as shown in Supplementary Fig.~\ref{Gxx}.
\begin{figure}[h!]
    \centering
    \includegraphics[width=0.95\linewidth]{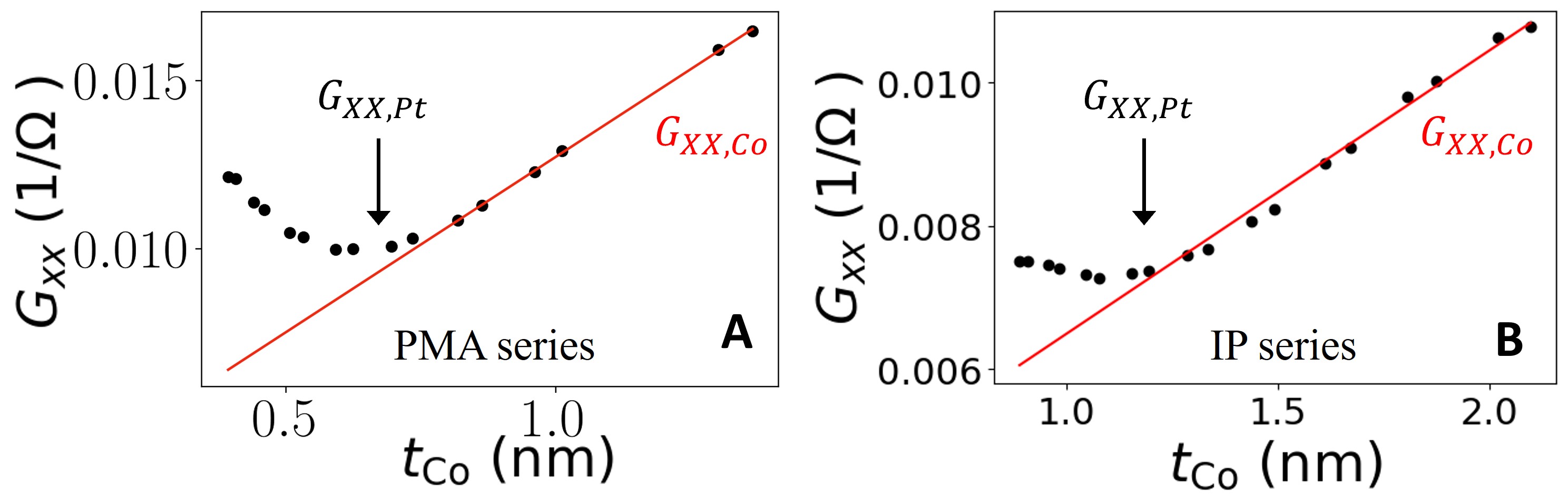}
    \caption{\textbf{Electrical conductance measurements.} Measured 4-point conductance of the fabricated devices as a function of the Co film thickness for \textbf{(A)} PMA series and \textbf{(B)} IP series. The red line is a fit to the linear regime of the data where a parallel resistor model of the HM/FM stack is appropriate.}
    \label{Gxx}
\end{figure}
In the very low Co thickness regime ($\sim$ 0.4 nm), the Co likely does not yet form a continuous film on top of the Pt so we expect the conductance measured here is entirely due to that of a bare Pt. As the Co thickness is increased, the conductance decreases due to increased surface scattering of conduction electrons in the Pt from the growing Co layer. In the regime above 0.8 nm of Co, the conductance is linear in the Co thickness, which is the expected behavior of a simple parallel-resistor model. We fit a line to the linear regime, the slope of which is the (inverse) resistivity of Co: 9.59 \textmu ohms cm for the PMA series and 25.26 \textmu ohms cm for the IP series. We estimate the resistivity of the 4 nm Pt layer adjacent to an established Co layer as corresponding approximately to the conductivity value at the minimum of the $G_{XX}$ (as indicated in the figure): 40 \textmu ohms cm for PMA series and 53.9 \textmu ohms cm for IP series. All of the Pt/Co/MgO samples for which we performed measurements of current-induced torque have Co layers thicker than 0.8 nm.

\subsection{Magnetometry}

To find the saturation magnetization  $M_s$ of the Co, we measure the magnetic moment on 3 mm $\times$ 3 mm thin films diced from the wafer adjacent to the patterned devices with vibrating sample magnetometry (VSM).  The magnetic moment it measures is ($\mu_0 M_s\text{Volume}$). If we divide this by area, we get $\mu_0 M_s t_\text{Co}$, which is an expression we use in the main text Eq.~\ref{taudef}. To get this quantity for each device, we plot it versus $t_\text{Co}$. $t_\text{Co}$ and linearly interpolate with the line shown in Supplementary Fig.~\ref{MsCo}.

\begin{figure}[h!]
    \centering
    \includegraphics[width=0.5\linewidth]{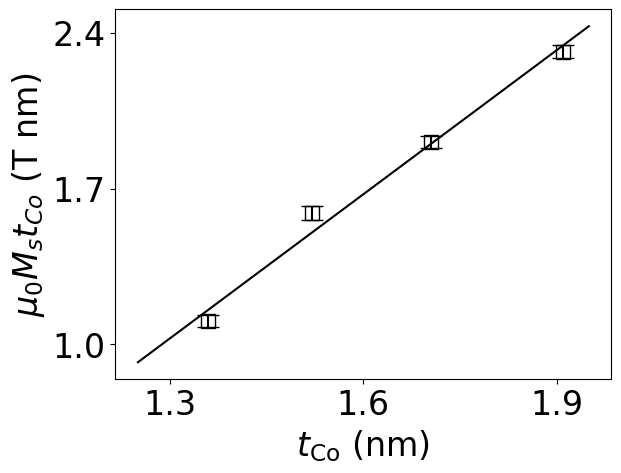}
    \caption{\textbf{Saturation magnetization measurements.} Measured saturation magnetization per unit area $\mu_0 M_s t_\text{Co}$ as a function of Co thickness $t_\text{Co}$. }
    \label{MsCo}
\end{figure}
\newpage

\subsection{Calibration of effective magnetization of samples with in plane anisotropy via spin-torque ferromagnetic resonance (ST-FMR) measurements}

We adopt the conventional ST-FMR measurements to obtain the effective magnetization $M_\text{eff}$ for devices with in-plane magnetic anisotropy. $M_\text{eff}$ is obtained by fitting the resonance peak using the Kittel formula\cite{Kittel1948}. We measured $\mu_0 M_\text{eff}$ for eight devices with varying $t_\text{Co}$ values, as shown below marked in black dots. We then fit the $\mu_0 M_\text{eff}$ versus $t_\text{Co}$ using a second order polynomial, and use the fitted curve to interpolate $\mu_0 M_\text{eff}$ for every device with in-plane anisotropy measured in main text Fig.~(\ref{torques}).
\begin{figure}[h!]
    \centering
    \includegraphics[width=0.5\linewidth]{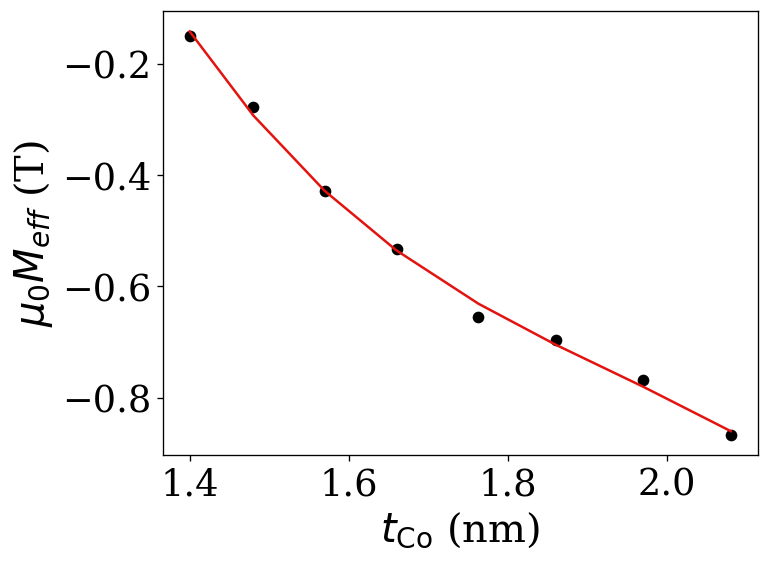}
    \caption{Effective magnetization measured using conventional ST-FMR for samples with in-plane magnetic anisotropy.}
    \label{Meff_STFMR}
\end{figure}


\newpage
\end{document}